\documentclass[12pt]{article}

\def\dalemb#1#2{{\vbox{\hrule height .#2pt
        \hbox{\vrule width.#2pt height#1pt \kern#1pt
                \vrule width.#2pt}
        \hrule height.#2pt}}}

\let\a=\alpha \let\b=\beta \let\g=\gamma \let\d=\delta \let\e=\epsilon
\let\z=\zeta  \let\th=\theta  \let\k=\kappa
\let\l=\lambda \let\m=\mu \let\n=\nu \let\x=\xi  
\let\s=\sigma \let\t=\tau    
 
      \let\G=\Gamma  \let\Th=\Theta \let\L=\Lambda
\let\X=\Xi  \let\S=\Sigma  \let\Y=\Psi
 
\let\la=\label  
  
\def\nn{\nonumber} \def\bd{\begin{document}} \def\ed{\end{document}}
\def\ds{\documentstyle} \let\fr=\frac \let\bl=\bigl \let\br=\bigr
\let\Br=\Bigr \let\Bl=\Bigl
\let\bm=\bibitem
\let\na=\nabla
\def\tU{{\widetilde U}}
\let\pa=\partial \let\ov=\overline
\def\ie{{\it i.e.\ }}
\newcommand{\be}{\begin{equation}}
\newcommand{\ee}{\end{equation}}
\def\ba{\begin{array}}
\def\ea{\end{array}}
\def\ft#1#2{{\textstyle{{\scriptstyle #1}\over {\scriptstyle #2}}}}
\def\fft#1#2{{#1 \over #2}}
\def\F#1#2{{ F_{#1}^{(#2)} }}
\def\cF#1#2{{ {\cal F}_{#1}^{(#2)} }}

\def\R{{\bf R}}
\def\sst#1{{\scriptscriptstyle #1}}
\def\oneone{\rlap 1\mkern4mu{\rm l}}
\def\e7{E_{7(+7)}}
\def\td{\tilde}
\def\wtd{\widetilde}
\def\im{{\rm i}}
\def\bog{Bogomol'nyi\ }
\newcommand{\ho}[1]{$\, ^{#1}$}
\newcommand{\hoch}[1]{$\, ^{#1}$}
\newcommand{\bea}{\begin{eqnarray}}
\newcommand{\eea}{\end{eqnarray}}
\newcommand{\ra}{\rightarrow}
\newcommand{\lra}{\longrightarrow}
\newcommand{\Lra}{\Leftrightarrow}
\newcommand{\ap}{\alpha^\prime}
\newcommand{\bp}{\tilde \beta^\prime}
\newcommand{\cB}{{\cal B}}
\newcommand{\cO}{{\cal O}}
\newcommand{\vecx}{\vec{x}}
\newcommand{\vecy}{\vec{y}}
\newcommand{\vecp}{\vec{p}}
\newcommand{\vecq}{\vec{q}}
\newcommand{\tr}{{\rm tr} }
\newcommand{\Tr}{{\rm Tr} }
\newcommand{\NP}{Nucl. Phys. }

\newcommand{\cL}{{\cal L}}
\newcommand{\cA}{{\cal A}}
\newcommand{\cD}{{\cal D}}

\def\sst#1{{\scriptscriptstyle #1}}
\def\0{{\sst{(0)}}}
\def\1{{\sst{(1)}}}
\def\2{{\sst{(2)}}}
\def\3{{\sst{(3)}}}
\def\4{{\sst{(4)}}}
\def\5{{\sst{(5)}}}
\def\6{{\sst{(6)}}}
\def\7{{\sst{(7)}}}
\def\8{{\sst{(8)}}}
\def\9{{\sst{(9)}}}
\def\ve{\varepsilon}
\def\vf{\varphi}
\def\F{\Phi}
\def\wg{\wedge}

\newcommand{\tamphys}{\it 
}

\newcommand{\auth}{AUTHORS}

\def\thb{\bar{\theta}}
\def\Thb{\bar{\Theta}}
\def\barp{\bar{p}}
\def\barq{\bar{q}}
\def\barc{\bar{c}}
\def\bard{\bar{d}}
\def\e{\epsilon}

\def \bi{\bibitem}
\def \la {\label}

\def \l {\lambda}
\def\foot{\footnote}
\def \tl  {{\tilde \l}}
\def \sql {{\sqrt \l}}
\def \adss {$AdS_5 \times S^5$\ }
\newcommand{\rf}[1]{(\ref{#1})}
\def \ov {\over}

\def\th{\theta}
\def\Th{\Theta}
\def\vth{\vartheta}
\def\btheta{{\bar\theta}}
\def\ttheta{{{\tilde\theta}}}
\def\bttheta{{{\bar\ttheta}}}
\def\vth{\vartheta}

\def\ra{\rightarrow}
\def\N{{\cal N}}
\def\F{{\cal F}}
\def\uM{\underline{M}}
\def\uN{\underline{N}}
\def\uP{\underline{P}}
\def\cc{\circ}
\def\eqv{\equiv}

\def\ni{\noindent}

\def\Ep{E^{{}^{(+)}}}
\def\Em{E^{{}^{(-)}}}

\def\Mp{M^{{}^{(+)}}}
\def\Mm{M^{{}^{(-)}}}

\def \ha{{1\ov 2}}

\def\r{\rho}

\def\Y{{\rm Y}}
\def\X{{\rm X}}
\def\tY{\tilde{\rm Y}}
\def\tX{\tilde{\rm X}}
\def\dY{\dot{\rm Y}}
\def\dX{\dot{\rm X}}

\def \J {\mathcal{J}}
\def \del {\partial}

\def\dF{\dot{F}}
\def\dG{\dot{G}}
\def\df{\dot{f}}
\def \E {{\cal E}}
\def \J {{\cal J}}

\def\ms{\mathcal{S}}
\def\mj{\mathcal{J}}
\def\soj{\fr{\ms}{\mj}}
\def \R {{\bf R}}
\def \om {\omega}
\def \bE {\bar E}
\def \x {{\cal X}}

\def \bi{\bibitem}
\def \la {\label}

\def \l {\lambda}
\def\foot{\footnote}
\def \tl  {{\tilde \l}}
\def \sql {{\sqrt \l}}
\def \adss {$AdS_5 \times S^5$\ }
\def \ov {\over}

\def \varpi {{\rm w}}

\def\thb{\bar{\theta}}
\def\Thb{\bar{\Theta}}
\def\zb{\bar{z}}
\def\psib{\bar{\psi}}
\def\barp{\bar{p}}
\def\barq{\bar{q}}
\def\barc{\bar{c}}
\def\bard{\bar{d}}
\def\e{\epsilon}
\def\wb{\bar{w}}
\def\lb{\bar{\l}}
\def\Jb{\bar{J}}
\def\Nb{\bar{N}}

\def\At{\tilde{A}}
\def\Bt{\tilde{B}}
\def\Ct{\tilde{C}}
\def\Dt{\tilde{D}}
\def\Et{\tilde{E}}
\def\Ft{\tilde{F}}
\def\Gt{\tilde{G}}
\def\Mt{\tilde{M}}
\def\at{\tilde{a}}
\def\bt{\tilde{b}}
\def\ct{\tilde{c}}
\def\dt{\tilde{d}}
\def\et{\tilde{e}}
\def\ft{\tilde{f}}
\def\gt{\tilde{g}}
\def\ola{\overleftarrow}
\def\ora{\overrightarrow}
\def\at{\tilde{\a}}

\def\ps{\rlap{\, /}\;\,p }
\def\ks{\rlap{\, /}\;\,k }

\def\gym{g_{YM}}

\def\adot{\dot{a}}
\def\bdot{\dot{b}}
\def\bpa{\bar{\pa}}

\begin{document}
\overfullrule=0pt
\parskip=2pt
\parindent=12pt
\headheight=0in \headsep=0in \topmargin=0in
\oddsidemargin=0in

\vspace{ -3cm}
\thispagestyle{empty}

\begin{center}

{\Large\bf Scattering of massive open strings in pure spinor
  }

 \vspace{.5cm} { I.Y. Park\footnote{Permanent address:
  Philander Smith College,
Little Rock, AR 72223, USA
 }  }
 \vskip 0.2cm

\vspace{0.5cm}
{\it Department of Physics, Pusan National University\\
Pusan 609-735, Korea \\
}
 \vspace{0.3cm}

\vspace{0.5cm}
{\it Physics Department and Center for Theoretical Physics\\
 Seoul National University\\
Seoul 151-742, Korea \\
}
 \vspace{0.3cm}



\vspace{0.5cm}
{\it Department of Physics and Research Institute of Basic Sciences\\
Kyung Hee University\\
Seoul 130-701, Korea \\
}


\end{center}

 \vspace{0.1cm}

 \begin{abstract}
\ni In {\em Phys.\ Lett.} {\bf B 660}, 583 (2008), it was proposed
that the D-brane geometry could be produced by open string quantum
effects. In an effort to verify the proposal, we consider scattering
amplitudes involving {\em massive} open superstrings. The main goal
of this paper is to set the ground for two-loop ``renormalization"
of an oriented open superstring on a D-brane and to strengthen our
skill in the pure spinor formulation of a superstring, an effective
tool for multi-loop string diagrams. We start by reviewing
scattering amplitudes of massless states in the 2D component method
of the NSR formulation. A few examples of massive string scattering
are worked out. The NSR results are then reproduced in the pure
spinor formulation. We compute the amplitudes using the unintegrated form of the massive
vertex operator constructed by Berkovits and Chandia in {\em JHEP}
{\bf 0208}, 040 (2002). We point out that it may be possible to
discover new Riemann type identities involving Jacobi
$\vartheta$-functions by comparing a NSR computation and the
corresponding pure spinor computation.

\end{abstract}
\newpage

\setcounter{equation}{0}
\setcounter{footnote}{0}
\setcounter{section}{0}


\section{Introduction}

To some extent, a D-brane \cite{Polchinski:1995mt}\cite{pol}\cite{Johnson:2000ch} is a peculiarity that originates from the endpoints of an open string.
Before the birth of D-brane physics, an oriented open string was viewed as inconsistent because it was believed impossible to consistently couple an oriented open string to a closed string due to the different amounts of supersymmetry.
With the advent of D-brane physics, open string theory has come to enjoy
a more elevated status in string theory, providing a birthplace for the black hole entropy, matrix theories and AdS/CFT correspondence with its varieties. The ground breaking results just mentioned are related to the peculiarity one way or another. There may be additional new physics waiting to be discovered that is associated with the endpoints. As we describe below, the physics associated with open string divergences may be one such example.

It was long known that open string theory has divergences; they were
not taken seriously before since open string theory was believed
pathological anyway (for the reason stated above). Now with the
elevated status of an open string, the divergence issue must be
given proper consideration.\footnote{ Our initial motivation for considering open string divergences was to
ultimately derive AdS/CFT correspondence and its generalization from
the first principle. For that purpose, two ingredients seem
necessary. The first is the connection between geometry and open
string loop effects, and the main goal of the work of
\cite{Park:2007mc} and the subsequent papers was to establish this
connection. The other ingredient is an open string conversion into a
closed string in certain circumstances. There are pieces of evidence
for this \cite{Nielsen:1973qs}\cite{Gibbons:2000hf}\cite{Sen:2000kd}
\cite{Park:2001bm}. Once these ingredients are established, we
believe that a large portion of AdS/CFT should follow from applying
S-duality. Other aspects of AdS/CFT may be proven following the
approach of \cite{Kawai:2007ek} } In a series of papers
\cite{Park:2007mc}\cite{Park:2008sg}\cite{Park:2008fp}, the
divergence issue was initiated taking the case of a D3-brane. A
divergence removal procedure was proposed: the proposal may be
viewed as ``{\em renormalization}" of type IIB open superstring on a
D-brane. If the picture is correct, infinitely many counter terms
(or counter vertex operators to be precise) would be required, and
in that sense the procedure could be taken as ``renormalization" of
a non-renormalizable theory. As often believed in quantum field
theory, non-renormalizablity may not be an indication of a genuine
pathology but rather a signal of new physics. We believe it is the
case with the oriented open superstring. Although infinite in
number, the counter terms may appear in a controlled manner,
yielding a curved geometry when summed up. (This is in contrast to
quantum field theory situation where no analogous phenomenon
occurs.) In other words,
the new physics may be that the D-brane curved geometry is produced by the quantum effects of open strings that are hosted by the D-brane(s). At a technical level, the generation of the curved geometry may be revealed through the renormalization. \\

The status of the proposal is as follows.
In \cite{Park:2008fp}, one-loop divergence cancellation was carried out. The two-loop
has been partially checked in \cite{Park:2009ki}. Although the
results so far are consistent with the proposal up to and including
the two-loop, it is only the quadratic terms\footnote{The quadratic
terms that are referred to here are not those of the free action. They are the terms that come from expanding the curved geometry. They are
different from the free action by a sign as discussed in \cite{Park:2008fp}.} in the, so-called,
large $r_0$-expansion that have played a role in the divergence cancellation mechanism.
For the verification of the proposal, it is
desirable to come up with an example where the higher
order curvature terms do participate in the mechanism. With the hope to see the participation of the higher curvature terms, we went on to the three-loop amplitude of the massless vector states in \cite{Park:2010xwa}. Unlike the lower loop cases, at three-loop the amplitude itself has not been computed even in ten dimensions, i.e., the case of a spacetime filling brane.
The computation was carried out using the pure spinor formulation\footnote{See \cite{Aisaka:2002sd}\cite{Grassi:2004xr}\cite{Oda:2007ak} for a few related works.} \cite{Berkovits:2000fe}\cite{Berkovits:2002qx}
\cite{Berkovits:2005bt}\cite{Berkovits:2006bk} that was developed using ingredients of \cite{Siegel:1985xj}. \\

With the three-loop amplitudes within reach, one can proceed to the three-loop renormalization.
However, the three-loop analysis is bound to be
complex. In addition to this, it seems \cite{Park:2010xwa} that there is room for better understanding of the three-loop regulator of the pure spinor formulation. (See \cite{Grassi:2010ca} for a recent related discussion.) Could there be a case that does not require a three-loop analysis and may yet unravel the role of the higher order curvature terms in the proposed renormalization?
Scattering of {\em massive} string may be worth examining in this regard
and for a few other reasons. (We will focus on three-point scattering amplitudes of massive states at tree and one-loop levels.) First of all, the higher curvature terms might begin contributing at two-loop order (and on) for the scattering of massive states.\footnote{Scattering of some higher spin and/or massive states has been recently discussed in \cite{Polyakov:2010sk} and \cite{Haertl:2010ks} in the NSR formulation.} (A priori, they could contribute even at one-loop. However, as we will argue later, it is unlikely to be the case.) The second reason - which is not entirely independent of the first - is
that scattering of massive states might be associated with
some kind of {\em near-extremal} geometry.\footnote{We thank M. M. Sheikh-Jabbari for the discussion on this point.} This is an interesting possibility to investigate. Another reason for considering the massive case is to accumulate experience in the pure spinor techniques. In the long run, it is expected that various renormalization analyses of the massless cases at two-loop and three-loop orders will be carried out in the pure spinor formulation. Therefore it will be useful to strengthen our skills with
simpler exercises.
Whereas massive three-point amplitudes at tree level are relatively simple (at least for bosonic states), the algebra involved in the one-loop case is comparable to that of the massless vector four-point amplitude. To assure the correctness of the results, we compute some of the amplitudes in the NSR formulation first. In general, the NSR formulation is more effective than the pure spinor formulation at tree level. At one-loop, it is so when the number of the external legs are less than or equal to four. (More remarks below on this.) As a matter of fact, the entire section 2 is devoted to the NSR formulation. Although the results are standard, we could not find any literature where, e.g., the massless vector four-point amplitude was obtained in the 2D component (as opposed to 2D superspace) NSR formulation. In the review, we collect all the necessary identities concerning Jacobi $\vartheta$-functions and explicitly show how they can be used to simplify the intermediate expressions of amplitudes.
\\

The rest of the paper is organized as follows.
In section 2, we start by reviewing a few massless amplitudes in the NSR formulation: three-point amplitudes at tree level and the vector four-point amplitude at one-loop level. We then turn to amplitudes of massive open strings. We explicitly demonstrate use of several Riemann identities involving Jacobi $\vartheta$-functions. Subsequently we compute the tree and one-loop amplitudes
of the two massless vectors and one three-index antisymmetric
tensor. As well-known, the one-loop scattering of purely massless states have the same kinematic factor as the corresponding tree
amplitudes. We will see below that the same is true for amplitudes that include massive states: the
one-loop amplitudes involving the first excited states have the same kinematics factor as the corresponding tree amplitudes.\footnote{
It is likely to imply
that at one-loop we will not be able to see, even with the massive states, the
possible role of the higher order curvature terms in the proposed mechanism of renormalization.
}
In section 3, we reproduce the results of the three-point amplitudes in the
pure spinor formulation. Both the NS and Ramond states are considered. We compute the amplitudes using the unintegrated\footnote{Two-loop amplitudes require use of the integrated form of the vertex operator which is currently unavailable. We have made a substantial amout of efforts to obtain the intergrated form only to conclude that the task deserves a separate work because of the complexity.}  form of the massive
vertex operator constructed by Berkovits and Chandia in \cite{Berkovits:2002qx}. Compared with the NSR formulation, the tree computation in the
pure spinor formulation is more involved. This is also true for one-loop $n$-point amplitudes with $n\leq 4$. However, for a higher point amplitude, Riemann type identities involving product of five or more Jacobi $\vartheta$-functions summed over the spin structures will be required in the NSR computation. They do not appear to be known in mathematical literature. Since the pure spinor formulation (being a variation of the Green-Schwarz formulation) does not require summing over the spin structure, it suggests a possibility that new Riemann type identities may be discovered by comparing a NSR result and the corresponding pure spinor result. Furthermore, the pure spinor formulation will be more effective
in the higher loop computations, which will be needed for two-loop and three-loop open string renormalization in the near future.
In section 4, we conclude with a summary and future directions. Our conventions and some useful relations are given in Appendix A and B. Part of the computation in the pure spinor formulation is presented in Appendix C.
\section{ Scattering of massive states in NSR formulation}
In general, tree and one-loop amplitudes can be computed fairly
effectively in the NSR formulation. The three- and four- point
amplitudes at tree- and one- loop levels were computed long ago.
However, the results are scattered in the literature. For example,
the work of \cite{D'Hoker:1988ta} was carried out in 2D superspace and analyzed closed string scattering. For scattering of oriented open superstrings, we
could not find any NSR computation that is as explicit and
direct as the computation presented below. We have decided to
put together the results in the literature in a coherent manner for
future purposes. In the section below, we review the tree and one-loop
four-point amplitudes of the vector vertex operators. Then we compute
several amplitudes that involve the three-index antisymmetric
tensor of the first excited states. These amplitudes will be
reproduced by using the pure spinor formulation in section 3.
In the 2D component notation, a certain number of picture changing operators must be inserted. In a given amplitude, the number of picture changing operators, $n_{PCO}$, is given by
\bea
n_{PCO}=2g-2+n_B+\fr{n_F}{2} \label{pco}
\eea
where $n_B$ ($n_F$) is the number of bosons (fermions) inserted. The Mandelstam variables are defined
\bea
s=-(k_1+k_2)^2 \quad t=-(k_2+k_3)^2 \quad u=-(k_1+k_3)^2
\eea
The massless vector vertex operators in the (-1)- and zero- picture are given by
\bea
V_{A,(-1)}&=& e^{-\phi}\psi^\m e^{ik\cdot X}\nn\\
V_{A,(0)}&=& (i\pa X^\r+2\a' (k\cdot \psi) \psi^\r)
e^{ik\cdot X}
\eea
and the vertex operator for the three-index anti-symmetric tensor is given by \cite{Koh:1987hm} \cite{Tanii:1987bk}
\bea
V_{b,(-1)}&=& e^{-\phi}\,
(\psi^{\m}\psi^{\n}\psi^{\k}) e^{ik\cdot X}
\nn\\
V_{b,(0)}&=&
\Big[\fr{}{}i\pa X^{\m}
\psi^{\n}\psi^{\k}-i\pa X^{\n}\psi^{\m}\psi^{\k}
+i\pa X^{\k}\psi^{\m}\psi^{\n}
+(\a_0\cdot \psi)\psi^{\m}\psi^{\n}\psi^{\k}
\Big]e^{ik\cdot X}
\eea
\subsection{review of the massless case}
Below we start by reviewing the massless vector three-point amplitude.
The tree-level four-point amplitude can be computed similarly as explained, e.g., in \cite{pol}.
The one-loop diagrams in the NSR formulation is more complex because
of the involvement of various Riemann identities. After that, we present detailed steps of
the computation including all the required Riemann identities.
We use the 2D component notation and employ some of the results that were obtained in \cite{D'Hoker:1988ta} in the 2D superspace techniques.

\subsubsection*{massless vector three-point amplitude at tree-level}
Consider the three-vector scattering at the tree level, $<V_AV_AV_A>$. We loosely denote the amplitude by $<AAA>$. Eq.\rf{pco} yields $n_{PCO}=1$, and one possible choice of setup is to start with three picture (-1) operators and insert one PCO.
Going through the procedure that is explained, e.g., in (12.5.3) and (12.5.13) of \cite{pol}, one gets two (-1)-picture operators and one (0)-picture operator.
The tree level correlators can be computed based on the following two-point functions,
\bea
<X^\m(x) X^\n(y)>&=& -2\a'\eta^{\m\n} \ln|x-y| \nn\\
<\psi^{\m_1}(x)\psi^{\n_2}(y)>&=& \fr{\d^{\m_1\n_2}}{x-y}
\label{ss}
\eea
One can easily show
\begin{eqnarray}
&& <c(x_1)c(x_2)c(x_3)><e^{-\phi(x_1)}e^{-\phi(x_2)}>\nn\\
&& <\psi^\m(x_1)e^{ik_1\cdot X}\psi^\n(x_2)e^{ik_2\cdot X}
(i\pa X^\r+2\a' (k_3\cdot \psi) \psi^\r)e^{ik_3\cdot X}>\nn\\
\rightarrow && -\;\fr{x_{23}}{x_{12}}\;\eta^{\m\n}k_1^\r
-\;\fr{x_{13}}{x_{12}}\;\eta^{\m\n}k_2^\r
+\eta^{\m\r}k_3^\n-\eta^{\n\r}k_3^\m
\label{3ex1}
\end{eqnarray}
which, upon multiplying $\z_1^\m \z_2^\n \z_3^\r$, yields
\bea
(\z_1\cdot \z_2)(k_1\cdot \z_3)+
(\z_2\cdot \z_3)(k_2\cdot \z_1)+(\z_3\cdot \z_1)
(k_3\cdot \z_2)
\label{massless3}
\eea
Permutations change only the overall numerical coefficient, and eq.(\ref{massless3}) can be re-expressed up to an overall numerical factor as
\bea
\z_{1\m} \z_{2\n} \z_{3\r} V^{\m\n\r}
\eea
with
\[
V^{\m\n\r} \equiv \eta^{\m\n}(k_1^\r-k_2^\r)+\eta^{\n\r}(k_2^\m-k_3^\m)
+\eta^{\r\m}(k_3^\n-k_1^\n)
\]
The four-point tree
amplitude can be similarly computed. As a matter of fact, the NSR
computation is very similar to the computation in the Green-Schwarz
formulation where all the vertex operators (including the ones that correspond to a bra- and a ket- states) were treated on an equal footing \cite{Park:2008fp}.

\subsubsection*{massless vector four-point at one-loop}
The next example is massless vector four-point amplitude at one-loop.\footnote{Regarding this particular amplitude, the Green-Schwarz formulation may be the simplest. However, NSR formulation is better suited for higher n-point amplitudes or scattering of excited states as the number of inserted fields increases.
} The corresponding computation for closed string theory was done, e.g., in \cite{D'Hoker:1988ta} using 2D superspace techniques. To
make a connection with the literature such as \cite{pol}, we adopt here
the 2D component approach. The open string four-point amplitude can be
obtained by appropriately adjusting and tailoring some of the results in \cite{D'Hoker:1988ta}. Most of the results can be carried over to the open string analysis with minor modifications.
Imposing the boundary normal ordering as in \cite{pol}, one gets for the bosonic Green's function
\bea
G'(x,y;\t)=
-\a'\ln \left|\fr{\vartheta_1(x-y,\t)}{\vartheta'(0,\t)} \right|^2
-\a'\fr{\pi}{2\t}|x-y|^2, \label{2H31_quoted}
\eea
where the prime on $G$ indicates the absence of the zero modes and $\t$ is the modulus of annulus. $\vartheta$ is a Jacobi theta function:
\bea
\vartheta_1(x,\t)\equiv - \vartheta_{ab}(x,\t)\;\; \mbox{with} \;\;(a,b)=(1,1)
\eea
A few facts about Jacobi $\vartheta$-functions are summarized in Appendix A.
The fermionic two-point function is given by
\bea
S_{\n}(x,y)&=&<\psi^\m(x)\psi^\n(y)>_\n
=\eta^{\m\n} \fr{\vartheta[\n](x-y,\t)\vartheta_1'(0,\t)}
{\vartheta[\n](0,\t)\vartheta_1(x-y,\t)}
\eea
where the subscript $\n$ represents the spin structure. For
one-loop, there are four structures: ``even" structures,
$\n=(0,0),(0,1),(1,0)$ and an odd structure, $\n=(1,1)$. The four
vector scattering amplitude is given by
\bea
&& \z_1^{\m_1}\z_2^{\m_2} \z_3^{\m_3} \z_4^{\m_4}\fr12\int
\fr{d\t}{2\t}
\sum_{\n}C_\n<(bc)\nn\\
&& \int \prod_{i=1}^4 dx_i (i\dot{X}^{\m_1}+2\a' k_1\cdot \psi \psi^{\m_1})e^{ik_1\cdot X(x_1)}
(i\dot{X}^{\m_2}+2\a' k_2\cdot \psi \psi^{\m_2})e^{ik_2\cdot X(x_2)} \nn\\
&& (i\dot{X}^{\m_3}+2\a' k_3\cdot \psi \psi^{\m_3})e^{ik_3\cdot X(x_3)}
(i\dot{X}^{\m_4}+2\a' k_4\cdot \psi \psi^{\m_4})e^{ik_4\cdot
X(x_4)}>_{\n} \label{pol4point}
\eea
One can take the coefficient, $C_\n$, as
$C_{1,0}=C_{0,1} =-C_{0,0}$ \cite{D'Hoker:1988ta}.\footnote{One need not be concerned with $C_{1,1}$: as well known, the odd structure only contributes to six- and higher- point amplitudes due to the presence of 2D fermionic zero-modes.}
Expansion of the matter part of the correlator in \rf{pol4point}
yields several types of terms. The types of terms that need to be computed are\footnote{The correlators with three $X$'s vanishes since there is an odd
number of $\psi$ fields.}
\[
<XXXX>, <XX(k\psi\psi)(k\psi\psi)>
\]
\[
<X(k\psi\psi)(k\psi\psi)(k\psi\psi)>,
<(k\psi\psi)(k\psi\psi)(k\psi\psi)(k\psi\psi)>
\]
The first three correlators vanish for various reasons.
\footnote{The fields with the same arguments do not get contracted by way of regularization. As far as we know, this should be understood as
dimensional regularization for the following reason. As two coordinates, $(x,y)$, approach each other, the Green's functions become the same as the corresponding tree-level Green's functions. At tree-level, a Green's function with the same arguments is omitted in dimensional regularization.}
The first term trivially vanishes due to a well-known identity,
\bea
\sum_\n C_\n \vartheta_{ab}(0,\t)^4=0
\eea
One can show straightforwardly that the second and the third terms vanish as well due to the Riemann
identities, \rf{3ptriemann}. The expected one-loop result should come solely from the fourth correlator,
\bea
<(k\psi\psi)(k\psi\psi)(k\psi\psi)(k\psi\psi)>
\eea
Applying the standard Wick contractions produces various terms, which then get multiplied with the polarization vectors,
$\z_1^{\m_1}\z_2^{\m_2} \z_3^{\m_3} \z_4^{\m_4} $, in front.
Let us discuss a few examples. After some algebra, one can show that the coefficient of $(\z_1\cdot \z_2)(\z_3\cdot \z_4)$
is
\bea
&&\fr14\;\sum_{\n}C_\n \vartheta_{ab}(0,\t)^4
\Big[ t^2 S_\n(x_1-x_2)S_\n(x_1-x_4)S_\n(x_2-x_3)S_\n(x_3-x_4)\nn\\
&&\hspace{1.3in}- u^2 S_\n(x_1-x_2)S_\n(x_1-x_3)S_\n(x_2-x_4)S_\n(x_3-x_4)\nn\\
&&\hspace{1.3in}+ s^2 S_\n(x_1-x_2)^2 S_\n(x_3-x_4)^2
\Big]e^{k_i\cdot k_j \ln \F(x_i,x_j)}
\label{z1z2z3z4}
\eea
where $\F$ is related to (\ref{2H31_quoted}) by
\bea
\F=\ln [-G'(x,y;\t)]
\eea
In (\ref{z1z2z3z4}), only the relevant factors among the factors present in \rf{pol4point} have been recorded.
The factor $\vartheta_{ab}(0,\t)^4$ arises as a result of evaluating part of the path integral as explained in Appendix A. The part in the square bracket results from the $\psi$-contractions.
Note that the sum $\sum_\n$ in (\ref{z1z2z3z4}) is over the even spin structures. The Riemann identity \rf{ri} with the fact that $\vartheta_{11}(0,\t)=0$ leads to
\bea
&&\sum_{\n=even}C_\n \vartheta_{ab}(0,\t)^4 S_\n(x_1,x_2)^2
S_\n(x_3,x_4)^2=\vartheta_1'(0,\t)^4 \nn\\
&&\sum_{\n=even}C_\n \vartheta_{ab}(0,\t)^4 S_\n(x_1,x_2)
S_\n(x_3,x_4)S_\n(x_1,x_3) S_\n(x_2,x_4)=\vartheta_1'(0,\t)^4\nn\\
&&\sum_{\n=even}C_\n \vartheta_{ab}(0,\t)^4
S_\n(x_1,x_2) S_\n(x_3,x_4)S_\n(x_1,x_4)
S_\n(x_2,x_3)=\vartheta_1'(0,\t)^4
\label{4ptriemann}
\eea
Using these in (\ref{z1z2z3z4}), one gets for the coefficient of
$\z_1\cdot \z_2\; \z_3\cdot \z_4$
\bea
\fr14\;\Big[(s^2-u^2+t^2)
\eea
The coefficients of $\z_1\cdot \z_3\; \z_2\cdot \z_4$ and
$\z_1\cdot \z_4\; \z_2\cdot \z_3$ can be similarly computed: putting them together, one gets
\bea
&&\fr14\;\Big[(s^2-u^2+t^2)\z_1\cdot \z_2\, \z_3\cdot \z_4
+(u^2-s^2-t^2)\z_1\cdot \z_3\, \z_2\cdot \z_4\nn\\
&&+(t^2-u^2+s^2)\z_1\cdot \z_4\, \z_2\cdot \z_3
\Big]e^{k_i\cdot k_j \ln \F(x_i,x_j)}
\label{v4semifinal}
\eea
For the final forms of the coefficients, permutations of the equation above must be taken into account.
Once permutations\footnote{To keep the factor $e^{k_i\cdot k_j \ln F(x_i,x_j)}$
out, the dummy indices $x_i$'s should be permuted in the same manner.} are added, eq.\rf{v4semifinal} yields the expected expression,
\bea
&&\fr12\;\Big[tu\; \z_1\cdot \z_2\, \z_3\cdot \z_4
+st\;\z_1\cdot \z_3\, \z_2\cdot \z_4+su\;\z_1\cdot \z_4\, \z_2\cdot \z_3
\Big]e^{k_i\cdot k_j \ln \F(x_i,x_j)}
\eea
For the second example,
let us consider the $(\z\cdot \z)\, (\z \cdot k)\,(\z\cdot k )$ type
terms and work out $(\z_1\cdot \z_2)\, (\z \cdot k)\,(\z\cdot k )$ to
be specific. Some of the terms vanish due to the Riemann
identities in (\ref{3ptriemann}). The coefficient of $(\z_1\cdot \z_2)$ turns out to be
\bea
-\vartheta_{ab}(0,\t)^4\Big[
-\fr{s}2\,\Big( \z_3\cdot k_4\, \z_4\cdot k_3\Big)
S_\n(x_1,x_2)^2S_\n(x_3,x_4)^2
\eea
\[
+ \Big(-\fr{s}{2}\;\z_3\cdot k_1 \z_4\cdot k_2
+\fr{u}{2}\;\z_3\cdot k_1 \z_4\cdot k_3
+\fr{u}{2}\;\z_3\cdot k_4 \z_4\cdot k_2\Big)
S_\n(x_1,x_2)S_\n(x_1,x_3)S_\n(x_2,x_4)
S_\n(x_3,x_4)
\]
\[
\;\;+\Big(\fr{s}{2}\;\z_3\cdot k_2\, \z_4\cdot k_1
-\fr{t}{2}\;\z_3\cdot k_2\, \z_4\cdot k_3
-\fr{t}{2}\;\z_3\cdot k_4\, \z_4\cdot k_1\Big)
S_\n(x_1,x_2)S_\n(x_2,x_3)S_\n(x_1,x_4)
S_\n(x_3,x_4)
\Big]
\]
This result is in terms of $S_\n$ but it can be re-expressed
in terms of $\vartheta_\n$
by using the Riemann identity, \rf{ri}
\bea
&&-\vartheta_1'(0,\t)^4\label{z1z2} \Big[
\Big(-\fr{s}{2}\;\z_3\cdot k_1 \z_4\cdot k_2
+\fr{u}{2}\;\z_3\cdot k_1 \z_4\cdot k_3
+\fr{u}{2}\;\z_3\cdot k_4 \z_4\cdot k_2\Big)\\
&&\hspace{.6in}+\Big(\fr{s}{2}\;\z_3\cdot k_2\, \z_4\cdot k_1
-\fr{t}{2}\;\z_3\cdot k_2\, \z_4\cdot k_3
-\fr{t}{2}\;\z_3\cdot k_4\, \z_4\cdot k_1\Big)
-\fr{s}2\,\Big( \z_3\cdot k_4\, \z_4\cdot k_3\Big)
\Big]\nn
\eea
There are five more terms of this type with different $\z_i \cdot \z_j$ in
front. Taking the permutations into account and simplifying
the resulting expression with the momentum
conservation and transversality of the polarization vectors, one gets
for the coefficient of $\z_1\cdot \z_2 $
\bea
-\fr12\, [\;t\, \z_3\cdot k_1\, \z_4\cdot k_2
+u\, \z_3\cdot k_2\, \z_4\cdot k_1]
\eea
which is precisely the expected result as can be seen by comparing with \rf{K}. Collecting the results so far, one gets up to an overall numerical factor\footnote{The additional factor of $\fr{1}{\t^4}$ as compared with \rf{pol4point} appears as a result of performing integration over $X$-zero modes as explained in \cite{D'Hoker:1988ta}. }
\bea
&&(2\pi)^{10}\d(\S_i k_i)K\int \fr{d \t}{\t^5}\int dx_1..dx_4 \nn\\
&& \left[\F(x_1,x_2)\F(x_3,x_4)\right]^{k_1\cdot k_2}
\left[\F(x_1,x_3)\F(x_2,x_4)\right]^{k_1\cdot k_3}
\left[\F(x_1,x_4)\F(x_2,x_3)\right]^{k_1\cdot k_4}
\eea
where the expression for kinematic factor, ${K}$, can be found in \rf{K}.
\subsection{scattering of massive states}
With the review in the previous section, the amplitudes involving massive states can be tackled.\footnote{In the scattering of massless states, the one-loop kinematic factors turned out to coincide with those of the corresponding tree amplitudes.
(For example, in the four vector scattering both the tree amplitude and the one-loop amplitude include the factor commonly called ${ K}$.)
In fact, using the 2D superspace notations of \cite{D'Hoker:1988ta} there is an indirect way of arguing that the bosonic coincidence (just mentioned) guarantees the coincidence in the 2D supersymmetric case, and the argument remains valid for the amplitudes involving massive states. The computations below explicitly confirm that the one-loop
amplitudes have the same kinematic factors as those of the corresponding tree
level amplitudes.} In this section, we will work out a few examples.
In section 3, we reproduce all the tree level amplitudes and the one-loop $<AAb>$ amplitude among the amplitudes computed here.
\subsubsection*{tree-level amplitudes of massive states}
For convenience, we choose the locations of the vertex operators as
\bea
x_1\ra
\infty, x_2=1, x_3=0
\eea
The mass of the tensor state is given by
\bea
k^2=-\fr1{2\a'}
\eea
For our first example, we consider $<V_AV_AV_b>$, which we loosely call
$<AAb>$.\footnote{As for the $<AAb>$ amplitude, the final form of the kinematic factor (i.e., $\z_1^{\m}\z_2^{\n}k_2^\r e_{3\m\n\r}$ in \rf{AAbNSR}) can be easily determined by momentum conservation and transversality of polarization tensors.} One may take the following equation as a starting point:
\[
\z_{\m_1}\z_{\m_2}e_{3\m_3\n_3\k_3}
<c(x_1)c(x_2)c(x_3)><e^{-\phi(x_1)}e^{-\phi(x_2)}>
\]
\[
<\Big(\psi^{\m_1}\Big)\Big(\psi^{\m_2}\Big)
\Big[\fr{}{}i\pa X^{\m_3}
\psi^{\n_3}\psi^{\k_3}-i\pa X^{\n_3}\psi^{\m_3}\psi^{\k_3}
+i\pa X^{\k_3}\psi^{\m_3}\psi^{\n_3}
+(\a_0\cdot \psi)\psi^{\m_3}\psi^{\n_3}\psi^{\k_3}
\Big]>
\]
\bea
= -6\;\z_1^{\m}\z_2^{\n}k_2^\r e_{3\m\n\r}
\label{AAbNSR}
\eea
To this, one should add $(1\leftrightarrow 2)$ contributions, which doubles the result.
For the second example, consider $<Abb>$ amplitude. After some algebra one gets
\begin{eqnarray}
&& \z_{1\m_1}e_{2\m_2\n_2\k_2}e_{3\m_3\n_3\k_3}
<c(x_1)c(x_2)c(x_3)><e^{-\phi(x_2)}e^{-\phi(x_3)}>\nn\\
&&<\left[\fr{}{}i\pa X^{\m_1}
+(\a_0\cdot \psi)\psi^{\m_1}
\fr{}{}\right]\left(\psi^{\m_2}\psi^{\n_2}\psi^{\k_2}\right)
\left(\psi^{\m_3}\psi^{\n_3}\psi^{\k_3}\right)>\nn\\
=&& -36 \;\z_{1}^\m e_{2}^{\m\r_2\s_2} e_{3}^{\k\r_2\s_2}k_1^\k
\end{eqnarray}
As above, the contribution from permutation $(2\leftrightarrow 3)$ should be added to this result. The final example of a three-point amplitude is $<bbb>$,
\[
e_{1\m_1\n_1\k_1} e_{2\m_2\n_2\k_2} e_{3\m_3\n_3\k_3}
<c(x_1)c(x_2)c(x_3)><e^{-\phi(x_1)}e^{-\phi(x_2)}>
<\left(\psi^{\m_1}\psi^{\n_1}\psi^{\k_1}\right)
\left(\psi^{\m_2}\psi^{\n_2}\psi^{\k_2}\right)
\]
\[
\left[\fr{}{}i\pa X^{\m_3}
\psi^{\n_3}\psi^{\k_3}-i\pa X^{\n_3}\psi^{\m_3}\psi^{\k_3}
+i\pa X^{\k_3}\psi^{\m_3}\psi^{\n_3}
+(\a_0\cdot \psi)\psi^{\m_3}\psi^{\n_3}\psi^{\k_3}
\fr{}{}\right]>
\]
The result turns out to be
\begin{eqnarray}
= -108\;e_1^{\r_1\m\n}e_2^{\r_2\m\n}e_3^{\r_1\r_2\k}k_2^\k
+96\;e_3^{\r_1\m\n}e_2^{\r_2\m\n}e_1^{\r_1\r_2\k}k_3^\k
-96\;e_1^{\r_1\m\n}e_3^{\r_2\m\n}e_2^{\r_1\r_2\k}k_3^\k
\end{eqnarray}
Taking permutations, $(k_2,e_2\leftrightarrow k_3,e_3)+(k_1,e_1\leftrightarrow k_3,e_3)$, into account, one gets
\begin{eqnarray}
<bbb>
= 300 \left( -\;e_1^{\r_1\m\n}e_2^{\r_2\m\n}e_3^{\r_1\r_2\k}k_2^\k
+\;e_3^{\r_1\m\n}e_2^{\r_2\m\n}e_1^{\r_1\r_2\k}k_3^\k
-\;e_1^{\r_1\m\n}e_3^{\r_2\m\n}e_2^{\r_1\r_2\k}k_3^\k
\right)
\end{eqnarray}

\subsubsection*{one-loop amplitudes involving massive states}
As well-known, a one-loop scattering amplitude of purely massless states
has the same kinematic factor as the corresponding tree
amplitude. We will see below that the same is true: the
one-loop amplitudes involving massive states have the same
kinematics factors as those of the corresponding tree amplitudes.
At one-loop, eq.\rf{pco} implies insertion of three picture changing operators: the AAb-amplitude at one-loop can be taken as
\bea
&& \z_1^{\m_1}\z_2^{\m_2} e_3^{\m_3\n_3\k_3}\fr12\int \fr{dt}{2t}
\sum_{\n}C_\n<(bc)\\
&& \int \Big(\prod_{i=1}^3 dx_i \Big) (i\dot{X}^{\m_1}+2\a' k_1\cdot \psi \psi^{\m_1})e^{ik_1\cdot X(x_1)}
(i\dot{X}^{\m_2}+2\a' k_2\cdot \psi \psi^{\m_2})e^{ik_2\cdot X(x_2)} \nn\\
&& (\fr{}{}i\pa X^{\m_3}
\psi^{\n_3}\psi^{\k_3}-i\pa X^{\n_3}\psi^{\m_3}\psi^{\k_3}
+i\pa X^{\k_3}\psi^{\m_3}\psi^{\n_3}
+(\a_0\cdot \psi)\psi^{\m_3}\psi^{\n_3}\psi^{\k_3})e^{ik_3\cdot X(x_3)}
>_{\n}\nn
\eea
Straightforward calculation yields
\bea
&& \z_{1\m_1}\z_{2\m_2}e_{\m_3\n_3\r_3}
<(i\dot{X}^{\m_1}+2\a' k_1\cdot \psi \psi^{\m_1})e^{ik_1\cdot X(x_1)}
(i\dot{X}^{\m_2}+2\a' k_2\cdot \psi \psi^{\m_2})e^{ik_2\cdot X(x_2)} \nn\\
&& (\fr{}{}i\pa X^{\m_3}
\psi^{\n_3}\psi^{\k_3}-i\pa X^{\n_3}\psi^{\m_3}\psi^{\k_3}
+i\pa X^{\k_3}\psi^{\m_3}\psi^{\n_3}
+(\a_0\cdot \psi)\psi^{\m_3}\psi^{\n_3}\psi^{\k_3})e^{ik_3\cdot X(x_3)}
>_{\n}\nn\\
=&& 6 k_3^2\,\z_1^{\m}\z_2^{\n}k_2^\r e_{3\m\n\r} S_\n(1,3)^2S_\n(2,3)^2
\left[\F(z_1,z_2)\right]^{k_1\cdot k_2}
\left[\F(z_1,z_3)\right]^{k_1\cdot k_3}
\left[\F(z_1,z_4)\right]^{k_1\cdot k_4}\label{AAB_massive}
\eea
Again, the function, $S_\n$, will not appear in the final expression. The reason is that \rf{AAB_massive}, multiplies by $\vartheta_{ab}(0,\t)^4$ and summed over the spin structures, is a special case of the first equation of the Riemann identities given in \rf{4ptriemann}.
Up to an overall numerical factor, one gets
\bea
&& (2\pi)^{10}\d(\S_i k_i)
( k_3^2\,\z_1^{\m}\z_2^{\n}k_2^\r e_{\m\n\r})
\int \fr{d \t}{\t^5}\int dx_1dx_2dx_3 \nn\\
&& \left[\F(x_1,x_2)\right]^{k_1\cdot k_2}
\left[\F(x_1,x_3)\right]^{k_1\cdot k_3}
\left[\F(x_1,x_4)\right]^{k_1\cdot k_4}
\eea
From now on, we will focus on the kinematics factors and not explicitly record the other factors such as $\F(x_i,x_j)$. Taking the spin structures into account , the $<Abb>$-amplitude is
\bea
&& \z_1^{\m_1} e_2^{\m_2\n_2\k_2}
e_3^{\m_3\n_3\k_3}\fr12\int \fr{dt}{2t}
\sum_{\n}C_\n<(bc)\; \int \Big(\prod_{i=1}^3 dx_i \Big)
(i\dot{X}^{\m_1}+2\a' k_1\cdot \psi \psi^{\m_1})e^{ik_1\cdot X(x_1)}
\nn\\
&& (\fr{}{}i\pa X^{\m_2}
\psi^{\n_2}\psi^{\k_2}-i\pa X^{\n_2}\psi^{\m_2}\psi^{\k_2}
+i\pa X^{\k_2}\psi^{\m_2}\psi^{\n_2}
+(\a_0\cdot \psi)\psi^{\m_2}\psi^{\n_2}\psi^{\k_2})e^{ik_2\cdot X(x_2)}
\nn\\
&& (\fr{}{}i\pa X^{\m_3}
\psi^{\n_3}\psi^{\k_3}-i\pa X^{\n_3}\psi^{\m_3}\psi^{\k_3}
+i\pa X^{\k_3}\psi^{\m_3}\psi^{\n_3}
+(\a_0\cdot \psi)\psi^{\m_3}\psi^{\n_3}\psi^{\k_3})e^{ik_3\cdot X(x_3)}
>_{\n}\nn\\
\eea
The types of terms that need further consideration are
$<(k_1 \psi \psi)(\psi \psi)(\psi \psi\psi \psi)>$ and
$<(k_1 \psi \psi)(\psi \psi\psi \psi)(\psi \psi\psi \psi)>$. (The other terms will vanish due to the odd numbers of the fermionic fields and/or dimensional regularization.) The latter turns out to vanish as we will show below but first consider the former.
One can show that the correlator, $<(k_1 \psi \psi)(\psi \psi)(\psi \psi\psi \psi)>$, yields
\bea
&& \z_1^{\m_1} e_2^{\m_2\n_2\k_2}
e_3^{\m_3\n_3\k_3}
( k_1\cdot \psi \psi^{\m_1})
( \pa X^{\m_2}\psi^{\n_2}\psi^{\k_2})
(k_3\cdot \psi\;\psi^{\m_3}\psi^{\n_3}\psi^{\k_3})
\nn\\
\Rightarrow && \pa_{z_2}G(z_2,z_1) \Big(-6 {\z_1}_{{\m}} {k_3}_{{\m}} {k_1}_{{\n}} {k_1}_{{\b}} {e_2}_{{\n}
{\r} {\a}} {e_3}_{{\r} {\a} {\b}}
+ 6 {\z_1}_{{\m}} {k_1}_{{\n}} {k_1}_{{\b}} {k_3}_{{\b}}
{e_3}_{{\m} {\r} {\a}} {e_2}_{{\n} {\r} {\a}}\Big)\nn\\
&&\!\!\! + \pa_{z_2}G(z_2,z_3) \Big( -6 {\z_1}_{{\m}} {k_3}_{{\m}} {k_3}_{{\n}} {k_1}_{{\r}} {e_2}_{{\n}
{\a} {\b}} {e_3}_{{\a} {\b} {\r}}
+6 {\z_1}_{{\m}} {k_3}_{{\n}} {k_1}_{{\r}} {k_3}_{{\r}}
{e_3}_{{\m} {\a} {\b}} {e_2}_{{\n} {\a} {\b}}\Big)
\label{ABB1}
\eea
where we have omitted the overall multiplicative factor, $S_\n(z_1,z_3)^2 S_\n(z_2,z_3)^2$. Summing \rf{ABB1} and the contribution from $(e_2,k_2)\Leftrightarrow (e_3,k_3)$ given by
\bea
&& \z_1^{\m_1} e_2^{\m_2\n_2\k_2}
e_3^{\m_3\n_3\k_3}
( k_1\cdot \psi \psi^{\m_1})
( \pa X^{\m_3}\psi^{\n_3}\psi^{\k_3})
(k_2\cdot \psi\;\psi^{\m_2}\psi^{\n_2}\psi^{\k_2})\nn\\
\Rightarrow &&
\pa_{z_2}G(z_2,z_1) \Big(-6 {\z_1}_{{\m}} {k_2}_{{\m}} {k_1}_{{\n}} {k_1}_{\k}
{e_2}_{{\n} {\a} {\b}} {e_3}_{{\a} {\b} \k}
+6 {\z_1}_{{\m}}
{k_1}_{{\b}} {k_2}_{{\b}} {k_1}_{\k} {e_2}_{{\m} {\n} {\a}}
{e_3}_{{\n} {\a} \k} \Big)\nn\\
&& +\pa_{z_2}G(z_2,z_3)\Big(
-6 {\z_1}_{{\m}} {k_2}_{{\m}} {k_1}_{{\n}} {k_2}_{\k}
{e_2}_{{\n} {\r} {\a}} {e_3}_{{\r} {\a} \k}
+6 {\z_1}_{{\m}}
{k_1}_{{\a}} {k_2}_{{\a}} {k_2}_{\k} {e_2}_{{\m} {\n} {\r}}
{e_3}_{{\n} {\r} \k}
\Big)
\label{ABB2}
\eea
one gets
\bea
&& \pa_{z_2}G(z_2,z_1) \Big(
6({k_1}\cdot {k_3}) {\z_1}_{{\m}} {k_1}_{{\k}}
{e_2}_{{\k} {\a} {\b}} {e_3}_{{\m} {\a} {\b}}
+6 ({k_1}\cdot {k_2}) {\z_1}_{{\m}}
{k_1}_{\k} {e_2}_{{\m} {\a} {\b}}
{e_3}_{\k{\a} {\b} } \Big)\;\; \nn\\
&&
+ \pa_{z_2}G(z_2,z_3) \Big(
6({k_1}\cdot {k_3}) {\z_1}_{{\m}} {k_3}_{{\k}}
{e_2}_{{\k} {\a} {\b}}{e_3}_{{\m} {\a} {\b}}
+6 ({k_1}\cdot {k_2}){\z_1}_{{\m}}{k_2}_{\k} {e_2}_{{\m} {\a} {\b}}
{e_3}_{{\a} {\b} \k}
\Big)\nn\\
\eea
Finally, Mathematica computation of $<(k_1 \psi \psi)(\psi \psi \psi \psi)(\psi \psi\psi \psi)>$ yields
\bea
&& \z_1^{\m_1} e_2^{\m_2\n_2\k_2}
e_3^{\m_3\n_3\k_3}
( k_1\cdot \psi \psi^{\m_1})
((k_2\cdot \psi)\psi^{\m_2}\psi^{\n_2}\psi^{\k_2})
((k_3\cdot \psi)\psi^{\m_3}\psi^{\n_3}\psi^{\k_3})\nn\\
\Rightarrow &&
-18k_3^2 \z_{1\m} e_{2\m\a\b} e_{3\a\b\n}k_{1\n}
+18k_2^2 \z_{1\m} e_{2\n\a\b} e_{3\a\b\m}k_{1\n}
+36 (\z_1\cdot k_2)e_{1\m} e_{2\m\a\b} e_{3\a\b\n}
k_{1\m}k_{1\n} \nn\\
\eea
The result vanishes once $(2\Leftrightarrow 3)$ contribution is added.

\section{Scattering of massive states in pure spinor}
In this section, we reproduce some of amplitudes computed in section 2 in the pure spinor formulation. We also compute several amplitudes that involve fermionic states. Our main goal is
to set the ground for the future work where the two-loop amplitudes
involving the first massive states will be computed. In the beginning we briefly review the non-minimal formulation. For the massive
vertex operator of the first excited states, only the unintegrated form is known \cite{Berkovits:2002qx}. It is necessary to know the form of the integrated vertex operator as well for a general amplitude. We will take construction of the integrated vertex operator elsewhere in the near future.\footnote{We have made preliminary attempts. Although the procedure
should be straightforward in principle, the steps seem to require some
intricate use of gamma matrix identities and the field equations.} With only the unintegrated vertex operator available, there are still a few amplitudes that can be computed, and they are the focus of the current section. As we will see, the superspace description of the first excited level introduces many auxiliary fields -which is typical in a superspace formulation: gauge fixing must be proceeded before amplitude computation. We discuss below that there is a natural gauge, and the theta-expansion will be implemented in that gauge.

\subsection{review of non-minimal formulation}
The non-minimal version of the pure spinor formulation contains several extra fields in addition to the usual string coordinates, $(X,\th)$: it contains the bosonic pure spinor fields, $(\l, \lb)$ with their canonical conjugates, $(w, \bar{w})$ and a constrained fermionic spinor, $r$ with its canonical conjugate, $s$.
Each field has different number of zero modes: for the bosonic
fields,
\bea
\begin{array}{cccccc}
X^m & \l^\a & w_\a & \lb_\a & \bar{w}^\a \\
10 & 11 & 11g & 11 & 11g \\
\end{array}
\label{nz}
\eea
and for the fermionic fields,
\[
\begin{array}{cccc}
\th^\a & d_\a & r_\a & s^\a \\
16 & 16g & 11 & 11g \\
\end{array}
\]
where $g$ denotes the number of loops.
Although the pure spinor formulation was formulated in the 16 component chiral notation, which is effective in most of the computations, we switch to the 32-component notation when more convenient. Manual manipulations become simpler and/or a new insight can be gained in some cases in the 32-component notation.
The relation between the 16 by 16 gamma matrices, $\g^m$, and the 32 by 32 gamma matrices, $\G^m$, is
\[
\G^m=\left(
\begin{array}{cc}
0 & (\g^m)_{\a\b} \\
(\g^m)^{\a\b} & 0 \\
\end{array}
\right)\;\;,\;\;
\]
The pure spinor constraint, $\l\g^m\l=0$, implies
\bea
(\l \g^m...)(... \g_m \l)=0
\eea
Defining the 32-component objects
\bea
\l_u=\left(
\begin{array}{c}
\l^\a \\
0 \\
\end{array}
\right),\quad
\l_d=\left(
\begin{array}{c}
0 \\
\l^\a \\
\end{array}
\right)
\eea
the constraint relation above translates into
\bea
(\l_u \G^m...)(...\G_m \l_d)=0 \label{lggl}
\eea
The following combination of $\l$-fields appears as a part of the
$[ds]$-integration measure
\bea
(\l\g_m)_{\k_1}(\l\g_n)_{\k_2}(\l\g_p)_{\k_3}\g^{mnp}_{\k_4\k_5}\;\;:
\;\;\mbox{ anti-symmetric in $\k$'s}
\label{anti-sym}
\eea
It is is totally antisymmetric in the $\k$-indices.
The basic OPEs \cite{Berkovits:2005bt} are
\bea
X^m(x)X^n(y)&=&-2\,\eta^{mn}\log|x-y|\nn\\
p_\a(x) \th^\b(y)&=& \fr{\d_\a^\b}{x-y} \label{pth}
\label{XXpth}
\eea
They lead to
\bea
&& d_\a \Pi_m\ra \fr{\g_{\a\b}^m}{y-z}\pa\th^\b,\;
\Pi^m V(z)\ra -\fr{2}{y-z}\;\fr{\pa}{\pa X^m}V(z)
\nn\\
&&d_\a(y) d_\b(z)\ra -\fr1{y-z}\g_{\a\b}^m \Pi_m(z),\;
d_\a V(z)\ra \fr{1}{y-z}D_\a V(z),\quad
\eea
where
\bea
d_\a &=& p_\a-\fr1{\a'} \g_{\a\b}^m\th^\b \pa X_m
-\fr1{4\a'} \g_{\a\b}^m {\g_m}_{\r\s}\th^\b \th^\r \pa\th^\s\nn\\
\Pi^m &=& \pa X^m+\fr12 \th \g^m \pa \th
\eea
and the covariant derivative is given by
 \bea
 D_\a = \fr{\pa}{\pa \th^\a}+\fr12\g_{\a\b}^m \th^\b \pa_m
 \label{covari}
 \eea
The OPEs between the currents are
\bea
&& N_{mn}(x)\l^\a(y)\ra \fr12\fr{(\g_{mn}\l)^\a}{x-y},\quad
J_\l(x)\l^\a(y)\ra \fr{\l^\a}{x-y} \nn\\
&& N^{kl}(x)N^{mn}(y)\ra -3\fr{\eta^{n[k}\eta^{l]m}}{(x-y)^2}
+\fr{\eta^{m[l}N^{k]n}-\eta^{n[l}N^{k]m}}{x-y}\nn\\
&& J_\l(x)J_\l(y)\ra -\fr{4}{(x-y)^2},\quad J_\l(x) N^{mn}(y)\ra
\mbox{regular}
\label{OPEs}
\eea
where
\bea
N_{mn}=\fr12 w\g_{mn}\l,\quad J_\l=w_\a \l^\a,\quad
\eea
Some of the OPEs above will be used below in the amplitude computation. The unintegrated and integrated forms of the massless vertex operator are given respectively by
\bea
V_A&=&\l^\a A_\a\nn\\
U_A&=&\pa\th^\a A_\a+\Pi^m A_m+d_\a W^\a+\fr12N^{mn}{\cal F}_{mn}
\label{U}
\eea
The unintegrated vertex operator for the first massive states was
obtained in \cite{Berkovits:2002qx}, and is given by
\bea
V_{B}=&& :\pa \th^\b \l^\a \g_{\a\b}^{mnp}B_{mnp}:
\nn\\
&&+\fr{1}{48}:d_\b \l^\a (\g^{mnpq})^\b{}_\a \pa_{[m}B_{npq]}:
+\fr37 :\Pi^m\l^\a (\g_{st}D)_\a B^{st}{}_m:\nn\\
&& +\fr{1}{16}:N^{mn}\l^\a \left( 3\,\pa_{[m}(\g_{|st|}D)_\a B^{st}{}_{n]}
-\fr{3}{7}\pa^q (\g_{q[m})^\b{}_\a (\g_{|st|}D)_\b B^{st}{}_{n]}\right):
\label{mvo}
\eea
The prescription for an arbitrary loop order has been written down. For our discussion we will need tree, one-loop and two-loop prescriptions. The prescription for computing the N-point tree amplitude is given
by
\bea
{\cal A}_{tree}=<{\cal N}_0(y)V_1(x_1)V_2(x_2)V_3(x_3)
\int dx_4 U_4(x_4)\cdots \int dx_N U_N(x_N)>
\eea
where
\bea
{\cal N}_0(y) 
=\exp(-\l(y)\bar{\l}(y)-r(y)\th(y))
\eea
is a tree-level regulator. As can be seen from above, the tree-level three-point amplitude requires only the unintegrated vertex operator.
The amplitudes prescriptions for the first two
loop orders are
\bea
{\cal A}_{1-loop}&=&\int d\t <{\cal N}_1(y)\int dw \m(w)b(w)V_1(x_1)
\int dx_2 U_2(x_2)\cdots \int dx_N U_N(x_N)>
\nn\\
{\cal A}_{2-loop}&=&\int d\t_1d\t_2d\t_3 <{\cal N}_2(y)
\prod_{s=1}^3\int dw_s \m(w_s)b(w_s)
\int dx_1 U_1(x_1)\cdots \int dx_N U_N(x_N)>
\nn\\
\label{loopprescription}
\eea
where the $b$ ghost is a composite field given by
\bea
b &=& s^\a\pa \lb_\a+\fr{\lb_\a\left[2\Pi^m(\g_m d)^\a
-N_{mn}(\g^{mn}\pa\th)^\a-J_\l\pa\th^\a
-\pa^2\th^\a\right]}{4\lb\l}
\eea
\bea
+\fr{(\lb\g^{mnp}\,r)(d\g^{mnp}d+24N_{mn}\Pi_p)}{192(\lb\l)^2}
-\fr{(r\g_{mnp}r)(\lb\g^md)N^{np}}{16(\lb\l)^3}
+\fr{(r\g_{mnp}r)(\lb\g^{pqr}r)N^{mn}N_{qr}}{128(\lb\l)^4} \label{bghost}
\nn
\eea
For one- and two- loops, one can use the regulator given in
\cite{Mafra:2009wq}
\bea
\N_{1,2}= e^{-\lb\l-r\th-\wb w+sd} \label{2loopreg}
\eea
The theta-expansions of the SYM fields were discussed in \cite{Harnad:1985bc}\cite{Ooguri:2000ps}\cite{Grassi:2004ih}\cite{Policastro:2006vt}.
In this work, we focus on amplitudes that involve the massless vector field (and the anti-symmetric three-index tensor field for the first excited states ):
the ten dimensional gaugino field is set to zero. For our computations, $\th$-expansion up to and including ${\cal O}(\th^5)$-order is required for some fields: we have
\bea
A_\a &=& \fr12 a_m (\g^m\th)_\a-\fr13 (\xi\g_m\th)(\g^m \th)_\a
-\fr1{32}F_{mn}(\g_p\th)_\a (\th\g^{mnp}\th)\nn\\
 &&+\frac{1}{60}(\g_m\th)_{\a} (\th \g^{mnp}\th)(\pa_n \xi\g_p\th)
-\fr1{576}(\g^m\th)_\a (\th \g_{m}{}^{sn}\th)(\th \g_{n}{}^{pq}\th)
\;\pa_s\pa_q a_p
+\cdots\nn\\
A_m &=& a_m-(\xi \g_m \th)-\fr18(\th\g_m\g^{pq}\th)F_{pq}
   +\frac{1}{12}(\th\g_m\g^{pq}\th)(\pa_n \xi \g_q\th)\nn\\  &&+\fr1{192} (\th \g_m{}^{st}\th)(\th \g_t{}^{pq}\th)
\pa_s F_{pq} +\cdots\nn\\
W^\a &=& \xi^\a -\fr14 (\g^{mn}\th)^\a F_{mn}
+\fr14(\g^{mn}\th)^\a F_{mn} +\fr1{48}(\g^{mn}\th)^\a(\th\g_n\g^{pq}\th)\pa_{m}F_{pq}\nn\\
&&-\fr1{96}(\g^{mn}\th)^\a(\th\g^{npq}\th)\pa_m\pa_p(\xi\g_q\th) -\fr1{1920}(\g^{mn}\th)^\a (\th \g_n{}^{st}\th)(\th \g_t{}^{pq}\th)
\pa_m \pa_s F_{pq}
+\cdots\nn\\
{\cal F}_{mn}&=& F_{mn}-2\pa_{[m}\xi \g_{n]}\th
+\fr14(\th\g_{[m}\g^{pq}\th)\pa_{n]}F_{pq}
  -\fr16 (\th\g_{[m}\g^{pq} \th)\pa_{n]}\pa_p (\xi\g_q \th)  \nn\\
&&-\fr1{96} (\th \g_{[m}{}^{st}\th)(\th \g^{tpq}\th)
\pa_{n]} \pa_s F_{pq}
+\cdots
\label{SYM}
\eea
The $\th$-expansion of the first excited states have not been written down in the literature. To implement the expansion, gauge fixing must be proceeded; we now turn to gauge fixing and the $\th$-expansion.

\subsection{$\th$-expansion of the massive vertex operator}
The unintegrated form of the vertex operator was obtained in
\cite{Berkovits:2002qx} but without its $\theta$-expansion. The
$\theta$-expansion of the superfield, $B_{mnp}$, that appears in the
expression for the massive vertex operator, \rf{mvo}, is essential for the amplitude computation. Below
we show that with a suitable gauge choice, $B_{mnp}$ can be put
into the following form:
\bea
B_{mnp}=b_{mnp}-2\,(\g_{[mn}\psi_{p]})_\k \th^\k
- \fr{1}{18}\;\g_{[\k_1\k_2}^q
\left(\g^{[mn}\right)_{\k_3]}{}^\d \pa_q\psi^{p]}_\d\,
\th^{\k_1}\th^{\k_2}\th^{\k_3}+O(\th^5)
\label{Bquoted}
\eea
Upon substituting in \rf{mvo}, one gets the massive vertex operator
in terms of $b_{mnp}$ and $\psi_\a^p$. For bosonic amplitudes, set $\psi=0$; one gets
for the vertex operator form of $B_{mnp}$
\bea
&& B_{mnp}=b_{mnp}+O(\th^5) \nn\\
\Rightarrow && B_{mnp} \equiv e^{mnp}\;e^{ik\cdot X}+O(\th^5)
\eea
where $e_{mnp}$ is a constant polarization tensor.
All the other terms in \rf{Bquoted} except the first term
seem to contain $\psi$ and its derivatives.
Although we have checked this to the order indicated,
it is likely that the full expression of $V_b$ is
\bea
B_{mnp}=e^{mnp}\;e^{ik\cdot X} \label{Vb}\;\;\mbox{when $\psi$ is set to zero}
\eea
For amplitudes that involve both the bosons and the fermions, one should keep $\psi$
as well in general:
\bea
B_{mnp} &=& \Big[e^{mnp}-2\,(\g_{[mn}\chi_{p]})_\k \th^\k
- \fr{i}{18}\;\g_{[\k_1\k_2}^q
\left(\g^{[mn}\right)_{\k_3]}{}^\d k_q\chi^{p]}_\d\,
\th^{\k_1}\th^{\k_2}\th^{\k_3}\Big]\;e^{ik\cdot X}+O(\th^5)
 \label{Vbpsi}\nn\\
\eea
where $\chi_\a^p$ is a constant wave function that satisfies $k^m \chi_{m\a}=0$. It is also constrained constrained by $\g_m^{\b\g}\chi_\g^m=0$.
We consdier a few examples of those types of amplitudes toward the end of subsection 3.3.

\vspace*{.2in}
\ni The $\th$-expansion can be implemented based on the results of
\cite{Berkovits:2002qx} and a gauge choice. It was stated around
eq.(5.3) of \cite{Berkovits:2002qx} that various field equations can be combined to
imply\footnote{In \cite{Berkovits:2002qx}, $[...]$ and $(...)$ were
defined without $\fr{1}{n!}$. We follow the same convention in this
subsection (i.e., section 3.2). In the most part of the next subsection, however, we use a convention where
\bea
[...]\equiv \fr{1}{n!} (\mbox{anti-symmetrization})
\eea
}
\bea
D_\a B^{mnp}=\g_{\a\b}^{[m}Z^{np]\b}-\fr1{48}(\g^{[mn})_\a{}^\b
H_\b^{p]}+\g_{\a\g}^{mnp}Y^\g \label{DBeq}
\eea
$H_\b^{p}$ is a
spin-3/2 superfield, and the precise identities of the superfields,
$Z^{np\b}$ and $Y^\g$ do not concern us.
By going to a special reference frame where the spatial momenta $k_a=0$ (the index, $a$, denotes the spatial
directions, $a=1,...,9$), the following relations were derived,
\bea
&& Z^{bc \g}=\fr14 (\g^{[b}\Psi^{c]})^\g,\quad H_\b^b=-72
\Psi_\b^b,\quad Z^{0b\g}=-\fr74\,(\g^{0}\Psi^{b})^\g,\quad \nn\\
&& H_\a^0=0,\quad B^{0bc}=0
\eea
The spin 3/2 superfiled, $\Psi_\g^c$
contains the physical spin 3/2 field, $\psi_\g^c=\Psi_\g^c|$, and is constrained by
$\g_c^{\b\g}\Psi_\g^c=0$. Upon substitution into \rf{DBeq}, these results imply
\bea
{D}_\b {B}^{abc}
&=&2\, (\g^{[ab}{\Psi}^{c]})_\b
\eea
which, by applying a Lorentz transformation to a generic frame
implies\footnote{ More explicitly, consider a Lorentz transformation
in the passive way. The LHS takes
\bea
{D'}_\b {B'}^{mnp}(X',\th')
&=& L_e^m L_f^n L_g^p\; S_\b^\r {D}_\r {B}^{efg}(X,\th)\nn\\
&=&2\, L_e^m L_f^n L_g^p\; S_\b^\r (\g^{[ef}\Psi^{g]})_\r(X,\th)\nn\\
&=&2\, (\g^{[mn}{\Psi'}^{p]})_\b(X',\th')
\eea
where $L, S$ denote the vector and spinor transformation matrices
respectively. In the first and the third equalities, the transformation
properties of fields have been used. }
\bea
{D}_\b {B}^{mnp}
&=&2\, (\g^{[mn}{\Psi}^{p]})_\b \label{DB}
\eea
The $\th$-expansions of $B_{mnp}$ and $\Psi_\g^c$ can be derived
from this by gauge fixing as follows. Consider
the $\th$-expanded forms of $B_{mnp}$ and $\Psi_\g^c$,
\bea
B^{mnp}&\equiv & B^{mnp}_\0+B^{mnp}_\1+B^{mnp}_\2+...
\equiv b^{mnp}+B^{mnp}_{\1\,\k}\;\th^\k
+B^{mnp}_{\2\,\k_1\k_2}\;\th^{\k_1}\th^{\k_2}+... \nn\\
\Psi_\g^c &\equiv & \Psi_{\0\g}^{c}+\Psi_{\1\g}^{c}+\Psi_{\2\g}^{c}+...
\equiv \psi_{\g}^{c}+\Psi_{\1\g\,\k}^{c}\;\th^\k
+\Psi_{\2\g\,\k_1\k_2}^{c}\;\th^{\k_1}\th^{\k_2}+...
\eea
Substituting these equations into (\ref{DB}), one gets
at the first two orders of the $\th$-expansion\footnote{The convention for the covariant derivative in \cite{Berkovits:2002qx} is
\bea
D_\a=\fr{\pa}{\pa\th^\a}+\g_{\a\b}^m \th^\b \pa_m
\eea
In this paper, we use the convention of, e.g.,\cite{Berkovits:2005bt} that is
quoted in \rf{covari}},
\bea
B_{\1\,\a}^{ mnp}&=&-2 (\g^{[mn}\Psi_\0^{p]})_\a\nn\\
2B_{\2\,\a\b}^{ mnp}&=&2 (\g^{[mn})_\a{}^\r \Psi_{\1\,\r\b}^{p]}
-\fr12\g_{\a\b}^q \pa_q B_\0^{mnp} \nn\\
-3B_{\3\,\a\k_1\k_2}^{ mnp}
&=&-\fr14 \g_{\a[\k_1}^q \pa_q B_{\1\,\k_2]}^{mnp}
+2 (\g^{[mn})_\a{}^\r \Psi_{\1\,\r\k_1\k_2}^{p]}
\label{Bcomp}
\eea
Using $\Psi_\0\equiv \psi$ and taking $[\a\b]$ and $[\a\k_1\k_2]$
parts in the second and the third equations respectively, one gets
\bea
B_{\1\,\a}^{ mnp}&=&-2 (\g^{[mn}\psi^{p]})_\a\nn\\
B_{\2\,\a\b}^{ mnp}&=& \fr12(\g^{[mn})_{[\a}{}^\r \Psi_{\1\,|\r|\b]}^{p]}
\nn\\
-3B_{\3\,\a\k_1\k_2}^{ mnp}
&=&-\fr1{12} \g_{[\a\k_1}^q \pa_q B_{\1\,\k_2]}^{mnp}
+\fr13 (\g^{[mn})_{[\a}{}^\r \Psi_{\1\,|\r|\k_1\k_2]}^{p]}
\label{Bcomp2}
\eea
The B-field has the usual gauge freedom, $B^{mnp}+\pa^{[m}
\L^{np]}$, that is associated with the field strength
\bea
C_{mnpq}\equiv \fr1{48}\pa_{[m}B_{npq]}
\eea
This freedom can be used to remove at least some of the auxiliary
fields. We illustrate this with $\Psi_\1$ that appears in the second and
third equations in \rf{Bcomp2}. As a matter of fact, $\Psi_\1$ would appear
in many equations that originate from comparing the higher order
$\th$-coefficients in \rf{DB}: could one set some part of $\Psi_\1$ or even the whole $\Psi_\1$? A desirable
gauge would be the one that retains the physical spectrum, $b_{mnp},
\psi_\g^c$ and the symmetric two-index tensor, $g^{mn}$, (and possibly their
derivatives). The fields, $b_{mnp}$ and $\psi_\g^c$, are the the
zeroth order components of $B_{mnp}$ and $\Psi_\g^c$ respectively.
As for the 44 bosonic degrees of freedom, $g^{mn}$, they are defined
to be the zeroth component of
\bea
G^{mn}=D\g^{(m} \Psi^{n)} \label{gmetric}
\eea
From the definition of $G^{mn}$ and the constraint
(\ref{gmetric}), one gets
\bea
g^{mn}\equiv G^{mn}|_{\th=0}=-\Tr [\g^{(m}\Psi_{\1}^{n)}]
\label{psigammatr}
\eea
This shows that $\Psi_{\1}^p$ contains $g^{mn}$: it is just
the gamma-trace part of $\Psi_\1$.
Therefore except the gamma trace part, \rf{psigammatr}, $\Psi_\1$
does not contribute to the physical content of the theory: we choose $\L_\2$ - which is the coefficient of the $\th$-quadratic term in the $\th$-expansion of the gauge parameter, $\L$ - appropriately such that
\bea
B_{\2\,\a\b}^{ mnp}&=& \fr12(\g^{[mn})_{[\a}{}^\r \Psi_{\1\,|\r|\b]}^{p]}
=0
\eea
Similarly it is not difficult to see that the $\Psi_\1$ part in the
third equation of \rf{Bcomp2} can be removed by adjusting $\L_\3$
appropriately: eq.\rf{Bcomp2} is now simplified as
\bea
B_{\1\,\a}^{ mnp}&=&-2 (\g^{[mn}\psi^{p]})_\a\nn\\
B_{\3\,\a\b\g}^{ mnp}
&=&-\fr1{18} \g_{[\a\b}^q (\g^{[mn}\pa_q\psi^{p]})_{\g]}
\eea
Note that $ B_\1$ and $B_\3$ are expressed in terms of $\psi$. Although
we checked this for the first few orders, it seems true that
one may gauge-fix $\L$ in
such a way that $B_{(n)}$ with $n\geq 1$ would be either zero or depend on $\psi$. In particular, this implies
\bea
B_{mnp}=b_{mnp}\;\; \mbox{when}\;\; \psi_\g^c=0
\eea
as indicated in \rf{Vb}.

\subsection{amplitudes involving massive states}
In this section, we compute a few examples of three-point amplitude
using the result of the previous section. We compute the
$<AAb>$\footnote{
One-loop amplitudes such as $<bbb>$ or $<Abb>$ have been computed in the previous section using the NSR formulation.
Reproduction of them in the pure spinor formulation must await the construction of the integrated vertex operators of the massive states.
} at the tree and one-loop level and confirm the results of the NSR analysis that the amplitude is proportional to the kinematic factor,\footnote{Using the momentum conservation and/or the transversality of the polarization tensors, this form can be rewritten in a few different forms. For example
\bea
\z_1^{\m} \z_2^{\n}k_2^\r\, e_{3\m\n\r}\,k_3^2=-2
\z_1^{\m} \z_2^{\n}k_2^\r\, e_{3\m\n\r}\,k_2\cdot k_3
\label{df2}
\eea
Due to the momentum conservation and the transversality of the
polarization tensors, these forms are the only forms that are
allowed when there are three factors of $k$'s. As we will see in the
next section, there are five factors of $k$'s in the case of
one-loop computation. In that case as well, the kinematic factor is
determined by the momentum conservation and the transversality of
the polarization tensors: $(k_3^2)^2$ appears instead of $k_3^2$. As
stated in the introduction, one of our main goals is to gain skills
through simple exercises and build an ``infrastructure" (such as
Mathematica programming) for more complicated computations. }
\bea
\z_1^{\m} \z_2^{\n}k_2^\r\, e_{3\m\n\r}\,k_3^2
\label{df}
\eea
Towards the end, we also compute a few amplitudes that involve fermions.

\subsubsection{bosonic amplitudes}

\subsubsection*{ $<AAb>$ tree amplitude}
As in the NSR analysis, we choose the locations of the vertex operators,
\bea
x_1\ra
\infty, x_2=1, x_3=0
\eea
Consider $<AAb>$ amplitude,
\bea
<(\l A)(\l A)V_B|_{\psi=0}>
\eea
where $V_B|_{\psi=0}$ denotes the antisymmetric three-index tensor part of $V_B$. For convenience, we quote the
vertex operator for the first excited states here again,
\bea
V_B|_{\psi=0}=&& (\l\g^{mnp}\pa \th)\, b_{mnp}
+\fr12 (d \g^{mnpq}\l) \pa_m b_{npq}
+\fr37\; \pa X^m (\l \g_{st}\g^l \th)\, \pa_l\,b^{st}{}_m\nn\\
&& +\fr18 N^{mn}\left[
3 (\l \g_{st}\g^l \th) \pa_m \pa_l \,b^{st}{}_n
+\fr37 (\l \g_{qm}\g_{st}\g^l \th)\, \pa^q \pa_l\, b^{st}{}_n
\right]
\eea
The first term in $V_b$ does not contribute due to the
absence of the $\th$-zero modes:
because the zero mode function does not depend on the 2D
coordinates, the term involving $\pa \th$ would
contribute only when $\pa \th$ gets contracted with another field. As can be seen by inspection, $\pa \th$ field does not get contracted with any other field.
Let us reexpress $V_B|_{\psi=0}$ as
\bea
V_B|_{\psi=0}=&& :\pa \th^\b \l^\a [C(X,\th)]_{\a\b}:
+:d_\b \l^\a [E(X,\th)]^\b{}_\a :
+ :\pa X^m\l^\a [F(X,\th)]_{m\a}:\nn\\
&& +:N^{mn}\l^\a [G(X,\th)]_{mn\a}:
\label{Vschematic}
\eea
where
\bea
C(X,\th)_{\a\b}&=& \g_{\a\b}^{mnp} \, b_{mnp} \nn\\
E(X,\th)^\b{}_\a &=& \fr12 ( \g^{mnpq})^\b{}_\a \pa_m b_{npq} \nn\\
F(X,\th)_{m\a} &=& \fr37\; ( \g_{st}\g^l \th)_\a\, \pa_l\,b^{st}{}_m \nn\\
G(X,\th)_{mn\a} &=& \fr18 \left[
3 ( \g_{st}\g^l \th)_\a \pa_m \pa_l \,b^{st}{}_n
+\fr37 ( \g_{qm}\g_{st}\g^l \th)_\a\, \pa^q \pa_l\, b^{st}{}_n
\right] \label{coeff}
\eea
We work out $ F(X,\th)$ contribution in detail in this subsection because it is the simplest among $E,F,G$. The computation of $F(X,\th)$ and $G(X,\th)$ is placed in Appendix C. After a few OPEs and some algebra, one can show\footnote{In the remainder of section 3.3 (and also in the appendices)
we use a convention where the anti-symmetrization has unit length:
\bea
[...]\equiv \fr{1}{n!} (\mbox{permutations})
\eea
}
\bea
&&<\l A^\1\,\l A^\2\,\Pi_m \l^\a F(X,\th)^\3_{m\a}>\\
&&= \fr{3i}{448}\Big(
\z_1^{m_1}\z_2^{n_2}e_3^{stm_3}k_2^{m_2}k_2^{m_3}k_3^{q_3}
<(\l\g^{m_1}\th)(\l\g^{p_2}\th)(\l\g_{st}\g^{q_3}\th)(\th\g_{m_2n_2p_2}\th)>\nn\\
&&\hspace{.5in}+ \z_2^{m_2}\z_1^{n_1}e_3^{stm_3}k_1^{m_1}k_2^{m_3}k_3^{q_3}
<(\l\g^{p_1}\th)(\l\g^{m_2}\th)(\l\g_{st}\g^{q_3}\th)(\th\g_{m_1n_1p_1}\th)>
\Big)\nn\\
&& \hspace*{4in} +(1\leftrightarrow 2)
\label{Fterm}
\eea
Using the identities given in \cite{Berkovits:2006bk} (quoted in Appendix B), one gets for the first two terms in \rf{Fterm}
\[
\z_1^{m_1}\z_2^{n_2}e_3^{stm_3}k_2^{m_2}k_2^{m_3}k_3^{q_3}
<(\l\g^{m_1}\th)(\l\g^{p_2}\th)(\l\g_{st}\g^{q_3}\th)(\th\g_{m_2n_2p_2}\th)>\nn\\
=-\fr{1}{180}\,\z_1^\m \z_2^\n k_2^\r\, e_{3\m\n\r}\, k_2\cdot k_3
\]
\[
\z_2^{m_2}\z_1^{n_1}e_3^{stm_3}k_1^{m_1}k_2^{m_3}k_3^{q_3}
<(\l\g^{p_1}\th)(\l\g^{m_2}\th)(\l\g_{st}\g^{q_3}\th)(\th\g_{m_1n_1p_1}\th)>\nn\\
= -\fr{1}{180}\z_1^\m \z_2^\n k_2^\r\, e_{3\m\n\r}\, k_1\cdot k_3
\]
The contributions coming from $(1\leftrightarrow 2)$ in
\rf{Fterm} simply doubles this result.

\subsubsection*{$<AAb>$ one-loop amplitude}
According to the one-loop prescription, the amplitude that we want to compute is
\bea
<(\l A)(\l A)V_B|_{\psi=0}>_{{1-loop}}&=&\int d\t <{\cal N}_{1}(y)\int dw \m(w)b(w)
\left(\int dx_1 U_A\right)\left(\int dx_2 U_A\right) V_b>
\nn\\
\eea
The number of zero modes is listed in \rf{nz}. At one-loop the amplitude has 16 $d$-zero modes.
To saturate the 16 $d$-zero modes, the only term in \rf{bghost} of the $b$ ghost that contributes is the term that contains $(d\g_{mnp}d)$; the only term of the massless vector vertex operators $U(1)$ ($U(2)$) that contributes is $[d_{\a_1}W^{\a_1}_\1]$ ( $[d_{\a_2}W^{\a_2}_\2]$); lastly the massive vertex operator contributes only through $[d_{\b} \l^{\a} E^{\b}{}_{\a}]$.
Collecting these, the relevant part of the one-loop amplitude is
\bea
{\cal K}&\equiv &\int [d\l][d\lb][dr][d\th][dw][d\wb][ds][dd]
e^{-\lb\l-r\th-\wb w+sd}
\nn\\
&& \fr{(\lb\g^{mnp}\,r)(d\g_{mnp}d)}
{192(\lb\l)^2}
[d_{\a_1}W^{\a_1}_\1]
[d_{\a_2}W^{\a_2}_\2][d_{\b_3} \l^{\a_3} E_\3^{\b_3}{}_{\a_3}(X,\th)]
\label{calK}
\eea
where $\doteq$ indicates that the overall numerical coefficient is not recorded precisely.
Carrying out $[ds]$ integration using the measure given in \cite{Gomez:2009qd}
\bea
[ds]&\doteq &\fr{1}{(\l\lb)^3}(\l\g^r)_{\a_1}(\l\g^s)_{\a_2}(\l\g^q)_{\a_3}
(\g_{rsq})_{\a_4\a_5}\e^{\a_1...\a_5\d_1...\d_{11}}
\pa^s_{\d_1}\cdots \pa^s_{\d_{11}}
\eea
one gets
\bea
{\cal K}&\doteq&\int [d\l][d\lb][dr][d\th][dw][d\wb][dd]
e^{-\lb\l-r\th-\wb w}
\nn\\
&&\fr{1}{(\l\lb)^3}(\l\g^r)_{\a_1}(\l\g^s)_{\a_2}(\l\g^q)_{\a_3}
(\g_{rsq})_{\a_4\a_5}\e^{\a_1...\a_5\d_1...\d_{11}}
d_{\d_1}\cdots d_{\d_{11}} \nn\\
&& \fr{(\lb\g^{mnp}\,r)(d\g_{mnp}d)}
{(\lb\l)^2}
[d_{\a_1}W^{\a_1}_\1]
[d_{\a_2}W^{\a_2}_\2][d_{\b_3} \l^{\r_3} E_\3^{\b_3}{}_{\r_3}(X,\th)]
\eea
Further integration over $d$ leads to
\bea
{\cal K}&\doteq&\int [d\l][d\lb][dr][d\th][dw][d\wb]
e^{-\lb\l-r\th-\wb w}
\nn\\
&& \fr{(\lb\g^{mnp}\,r)}
{(\lb\l)^5} (\l\g^r)_{\a_1}(\l\g^s)_{\a_2}(\l\g^q)_{\a_3}
(\g_{rsq})_{\a_4\a_5}\d_{\k_1...\k_5}^{\a_1...\a_5}
\nn\\
&&
[W^{\k_1}_\1]
[W^{\k_2}_\2][ \l^{\r_3} E_\3^{\k_3}{}_{\r_3}(X,\th)]
{(\g_{mnp})}^{\k_4\k_5}
\eea
It yields upon doing $r$-integration
\bea
&& {(\lb\g^{mnp}\,r)}
(\l\g^r)_{\a_1}(\l\g^s)_{\a_2}(\l\g^q)_{\a_3}
(\g_{rsq})_{\a_4\a_5}\d_{\k_1...\k_5}^{\a_1...\a_5}
\nn\\
&&
[W^{\k_1}_\1]
[W^{\k_2}_\2][ \l^{\r_3} E_\3^{\k_3}{}_{\r_3}(X,\th)]
{(\g_{mnp})}^{\k_4\k_5}\nn\\
\doteq && (\lb \g_{rsq} r) (\l \g^r W_\1)
(\l \g^s W_\2) (\l \g^q E_\3 \l)
\eea
The freedom mentioned below \rf{anti-sym} has been used to obtain the right-hand side. The field $r$ can be replaced by the covariant derivative $D$. The covariant derivative can act either on $W$'s or on $E_\3$. When it acts on the latter, the contribution can be dropped since the result is proportional to $\Psi_\g^c$ through the field equation and we are only considering the bosonic state setting $\Psi_\g^c=0$. The covariant derivative acting on $W$'s yields
\bea
\doteq
(\lb \g_{rsq} \g^{mn} \g^r \l){\cal F}_{\1 mn}(\l \g^q E_\3 \l)
(\l \g^s W_\2)-(\lb \g_{rsq} \g^{mn} \g^s \l)(\l \g^r W_\1){\cal F}_{\2 mn}
(\l \g^q E_\3 \l)\nn\\
\label{1loopmother}
\eea
The first term of (\ref{1loopmother}) yields\footnote{
Unlike the tree case, here there are five $k$-factors whereas there are only three factors of $k$'s in \rf{df}. It turns out that final results contain $(k_3^2)^2$ instead of $k_3^2$.(It must be the case because of the momentum conservation and transversality.) It is still compatible with \rf{df} since $k_3^2$ can be replaced by the on-shell value.
As in the NSR computation in section 2, the three-point amplitudes contain the factor
\bea
&& \left[F(x_1,x_2)\right]^{k_1\cdot k_2}
\left[F(x_1,x_3)\right]^{k_1\cdot k_3}
\left[F(x_1,x_4)\right]^{k_1\cdot k_4}
\eea
which is not explicitly recorded.
}
\bea
&&(\lb \g_{rsq} \g^{mn} \g^r \l){\cal F}_{\1 mn}(\l \g^q E_\3 \l)
(\l \g^s W_\2)\nn\\
=&& -\fr{i}{10}(\lb\l)k_1^m \z_1^nk_2^uk_2^s \z_2^q k_2^p
k_3^{\m_1}e_3^{\m_2\m_3\m_4} \Big[
(\l \g^{m\m_1\m_2\m_3\m_4} \l)(\l\g^{nuv}\th)
(\th \g^{vst}\th)(\th \g^{tpq}\th)\nn\\
&&\hspace{2in} +2\d_{[u}^n (\l \g^{m\m_1\m_2\m_3\m_4} \l)(\l\g_{v]}\th)
(\th \g^{vst}\th)(\th \g^{tpq}\th)
\Big]\nn\\
&& +{i}(\lb\l) k_1^s \z_1^t k_2^u k_2^p \z_2^q
k_3^{\m_1}e_3^{\m_2\m_3\m_4}\Big[
(\l \g^{m\m_1\m_2\m_3\m_4} \l)(\l\g^{nuv}\th)
k_1^{[n}(\th \g^{m]st}\th)(\th \g_{v}{}^{pq}\th)\nn\\
&&\hspace{2in} +2\d_{[u}^n (\l \g^{m\m_1\m_2\m_3\m_4} \l)(\l\g_{v]}\th)
k_1^{[n}(\th \g^{m]st}\th)(\th \g^{vpq}\th)
\Big]\nn\\
&& +\fr{i}2(\lb\l)k_1^s k_1^p \z_1^q k_2^u \z_2^v
k_3^{\m_1}e_3^{\m_2\m_3\m_4} \Big[
(\l \g^{m\m_1\m_2\m_3\m_4} \l)(\l\g^{nuv}\th)
k_1^{[n}(\th \g^{m]st}\th)(\th \g_{t}{}^{pq}\th)\nn\\
&&\hspace{2in} +2\d_{[u}^n (\l \g^{m\m_1\m_2\m_3\m_4} \l)(\l\g_{v]}\th)
k_1^{[n}(\th \g^{m]st}\th)(\th \g^{tpq}\th)
\Big]\nn\\
\label{1loop_1st}
\eea
As with the tree-level cases, manual evaluation of these terms is tedious; we rely on the
Mathematica package, Gamma.m \cite{Gran:2001yh}; it yields the following results:
\[
\d_{[u}^n (\l \g^{m\m_1\m_2\m_3\m_4} \l)(\l\g_{v]}\th)
(\th \g^{vst}\th)(\th \g^{tpq}\th)=0
\]
\[
k_1^{[n}k_1^s \z_1^t k_2^u k_2^p \z_2^q
k_3^{\m_1}e_3^{\m_2\m_3\m_4}
\d_{[u}^n (\l \g^{m\m_1\m_2\m_3\m_4} \l)(\l\g_{v]}\th)
(\th \g^{m]st}\th)(\th \g^{vpq}\th)
= \fr{1}{70} (k_3^2)^2 e_{3\m\n\r}k_1^\m \z_1^\n \z_2^\r
\]
\[
\d_{[u}^n (\l \g^{m\m_1\m_2\m_3\m_4} \l)(\l\g_{v]}\th)
(\th \g^{m]st}\th)(\th \g^{tpq}\th)=0
\]
\[
k_1^m \z_1^nk_2^uk_2^s \z_2^p k_2^q
k_3^{\m_1}e_3^{\m_2\m_3\m_4}
(\l \g^{m\m_1\m_2\m_3\m_4} \l)(\l\g^{nuv}\th)
(\th \g^{vst}\th)(\th \g^{tpq}\th)=0
\]
\[
k_1^{[n}k_1^s \z_1^t k_2^u k_2^p \z_2^q
k_3^{\m_1}e_3^{\m_2\m_3\m_4}
(\l \g^{m\m_1\m_2\m_3\m_4} \l)(\l\g^{nuv}\th)
(\th \g^{m]st}\th)(\th \g_{v}{}^{pq}\th)
=\fr{1}{105} (k_3^2)^2 e_{\m\n\r}k_1^\m e_1^\n e_2^\r
\]
\bea
&&k_1^{[n}k_1^s k_1^p \z_1^q k_2^u \z_2^v
k_3^{\m_1}e^{\m_2\m_3\m_4}
(\l \g^{m\m_1\m_2\m_3\m_4} \l)(\l\g^{nuv}\th)
(\th \g^{m]st}\th)(\th \g_{t}{}^{pq}\th)=0
\eea
These results confirm that the kinematic factor is equivalent
to \rf{df}.\footnote{The $(1\leftrightarrow 2)$ contribution simply doubles this result. }

\subsubsection{amplitudes involving fermions}

We now turn to amplitudes that involve fermionic states.
In general, the NSR formulation is most effective for a tree amplitude. For a demonstration, we first compute the scattering of two gauginos and a three index
tensor state. For the second example, we consider the amplitude of a vector and two spin 3/2 states.

\subsubsection*{$<\xi \xi b>$ tree and one-loop amplitudes}

As in the bosonic cases, one should collect, for the tree amplitude, the terms that have five $\th$'s among the terms that result by expanding $<(\l A)(\l A)V_B>$:
 \bea
 &&\Big[
 -\fr13 (\xi_1\g_{m_1}\th)(\l\g^{m_1} \th)
 +\frac{1}{60}(\l\g_{m_1}\th) (\th \g^{m_1n_1p_1}\th)(\pa_{n_1} \xi_1\g_{p_1}\th)
 \Big]\nn\\
 &&\Big[
 -\fr13 (\xi_2\g_{m_2}\th)(\l\g^{m_2} \th)
 +\frac{1}{60}(\l\g_{m_2}\th) (\th \g^{m_2n_2p_2}\th)(\pa_{n_2} \xi_2\g_{p_2}\th)
 \Big]\nn\\
 &&\Big[ (\l\g^{m_3n_3p_3}\pa \th)\, b_{m_3n_3p_3}
+\fr12 (d \g^{m_3n_3p_3q_3}\l) \pa_{m_3} b_{n_3p_3q_3}
+\fr37\; \pa X^{m_3} (\l \g_{s_3t_3}\g^{l_3} \th)\, \pa_{l_3}\,b^{s_3t_3}{}_{m_3}\nn\\
&& +\fr18 N^{m_3n_3}\Big(
3 (\l \g_{s_3t_3}\g^{l_3} \th) \pa_{m_3} \pa_{l_3} \,b^{s_3t_3}{}_{n_3}
+\fr37 (\l \g_{q_3m_3}\g_{s_3t_3}\g^{l_3} \th)\, \pa^{q_3} \pa_{l_3}\, b^{s_3t_3}{}_{n_3}
\Big)\Big] \label{xixib_tree}
\eea
While it is possible to directly evaluate all the five-$\th$ terms, it will be an extremely tedious task. One can save a large amount of algebra by noting that there are only a few possible kinematic factors that can be produced due to the momentum conservation, mass-shell conditions and constraints on the polarization vectors.
We illustrate the idea with one of the terms that one gets by expanding \rf{xixib_tree},
 \bea
 &&(k_2)_{n_2}(k_3)_{m_3}e_{n_3p_3q_3} (\xi_1\g_{m_1}\th)(\l\g^{m_1} \th)
 (\l\g_{m_2}\th) (\th \g^{m_2n_2p_2}\th)( \xi_2\g_{p_2}\th)
 (d \g^{m_3n_3p_3q_3}\l) \nn\\
 \doteq &&
 (k_2)_{n_2}(k_3)_{m_3}e_{n_3p_3q_3} (\xi_1\g_{m_1}\g^{abc}\g_{p_2}\xi_2)(\th \g^{abc}\th)(\l\g^{m_1} \th)
 (\l\g_{m_2}\th) (\th \g^{m_2n_2p_2}\th)
 (d \g^{m_3n_3p_3q_3}\l)
 \label{xixib_treesub}\nn\\
 \eea
where "$\doteq$" indicates that we have omitted an overall numerical factor. With this expression, one can carry out the operator product between $d$ and $\th$, which produces expressions that can be further evaluated using the identities given in Appendix B. At some point of the algebra and on, it is contractions $(k_2)_{n_2}(k_3)_{m_3}e_{n_3p_3q_3}$ and one of $(\xi_1 \g^{t_1} \xi_2), (\xi_1 \g^{t_1t_2t_3} \xi_2)$ and $(\xi_1 \g^{t_1...t_5} \xi_2)$. An inspection reveals
that the only potentially non-vanishing contraction is
 \bea
 (k_2\cdot k_3)e_{t_1t_2t_3}(\xi_1 \g^{t_1t_2t_3} \xi_2)\sim
  e_{t_1t_2t_3}(\xi_1 \g^{t_1t_2t_3} \xi_2)
 \eea
 where on the right hand side, momentum conservation was used and the on-shell value of $k_3^2$ has been omitted.
The presence of this term implies an unreasonable feature that the amplitude does not depend on the orientations of the momentum vectors. We have explicitly checked for  \rf{xixib_treesub} (and for some of the terms in \rf{xixib_tree}) that
such a term does not survive. A similar reasoning implies that the one-loop amplitude vanishes. The fact that the one-loop vanishes is also implied by what seems to be a general feature of one-loop amplitudes: one-loop kinematic factors are the same as those of the corresponding tree amplitudes.

\subsubsection*{$<A\psi\psi>$ tree amplitude}

\ni It is possible to compute $<A\psi\psi>$ tree amplitude without knowing the integrated form of the massive vertex operator. To compute it, we consider the form of the vertex operator in \rf{Vbpsi} with $e_{mnp}=0$.
(The first term in \rf{Vbpsi}, $e^{mnp}$, will not play a role here since we
consider $<A\psi\psi>$ amplitude. It would be relevant in an amplitude such as $<b\psi\psi>$.)
  Substituting \rf{Vbpsi} in \rf{mvo}, one gets
\bea
V_B|_{b=0}=&& :\pa \th^\b \l^\a [\tilde{C}(X,\th)]_{\a\b}:
+:d_\b \l^\a [\tilde{E}(X,\th)]^\b{}_\a :
+ :\Pi^m\l^\a [\tilde{F}(X,\th)]_{m\a}:\nn\\
&& +:N^{mn}\l^\a [\tilde{G}(X,\th)]_{mn\a}:
\label{Vfschematic}
\eea
where
\bea
\tilde{C}(X,\th)_{\a\b}&=&  \g_{\a\b}^{mnp}
            \Big[-12(\g_{mn}\chi_p)_\k \th^\k
        -2ik_q (\g^q)_{\k_1\k_2}(\g^{mn}\chi^p)_{\k_3}\;\th^{\k_1}\th^{\k_2}\th^{\k_3}   \Big]e^{ik\cdot X}\nn\\
\tilde{E}(X,\th)^\b{}_\a
                       &=& -6  ik^m(\g^{mnpq})^{\b}{}_\a
              (\g_{np}\chi_q)_\k \th^\k
                          e^{ik\cdot X} +{\cal O}(\th^5)\nn\\
\tilde{F}(X,\th)_{m\a}
        &=& \frac{18}{7} \Big[
     2(\g^{st}\g_{[st}\chi_{m]})_\a-\fr23 ik^q (\g_{st}\g^q)_{\a\k_2}
     (\g^{[st}\chi^{m]})_{\k_3} \;\th^{\k_2}\th^{\k_3}\nn\\
     &&\quad
       +2ik^n\, (\g^{st}\g^n)_{\a\b} (\g_{[st}\chi_{m]})_\k\, \th^{\b}\th^{\k}
         +{\cal O}(\th^6)\Big]\nn\\
\tilde{G}(X,\th)_{mn\a}
   &=&\fr94
   \Big\{
       ik^m\Big[2(\g^{st}\g_{[st}\chi_{m]})_\a+\fr23 ik^q (\g_{st}\g^q)_{\a\k_2}
            (\g^{[st}\chi^{m]})_{\k_3}\th^{\k_2}\th^{\k_3}\nn\\
            &&\quad
       +2ik^n\, (\g^{st}\g^n)_{\a\b} (\g_{[st}\chi_{m]})_\k\, \th^{\b}\th^{\k}
      +{\cal O}(\th^6) \Big]\nn\\
    &&+\frac{i}{7}k^p
         \Big[2(\g_{pm}\g^{st}\g_{[st}\chi_{n]})_\a+\fr23 ik^q (\g_{pm}\g_{st}\g^q)_{\a\k_2}
            (\g^{[st}\chi^{n]})_{\k_3}\th^{\k_2}\th^{\k_3}\nn\\
            &&\quad
       +ik^r\, (\g_{pm}\g^{st}\g^r)_{\a\b} (\g_{[st}\chi_{n]})_\k\, \th^{\b}\th^{\k}
     +{\cal O}(\th^6)  \Big]
   \Big\}
  \label{fcoeff_gen}
\eea
The wavefunction $\chi$ satisfies
 \bea
k^m \chi_{m\a}=0, \quad \g_m^{\b\g}\chi_\g^m=0
 \eea

\ni For the tree amplitude,
one should collect the terms that contain five $\th$'s out of the terms that result by expanding $<(\l A)(V_B|_{b=0})(V_B|_{b=0})>$. Some of the resulting terms vanish for obvious reasons, and can be easily omitted.
(For example, the $\Ct$ term can only appear with the $\Et$ term.)
The amount of the algebra involved is large even with the help of the Mathematica package Gamma.m: we will not attempt a full evaluation of $<(\l A)(V_B|_{b=0})(V_B|_{b=0})>$. For an illustration we have computed the term containing $\Ct \Et$  explicitly ; after lengthy and tedious manipulations, we have obtained
 \bea
 &&<(\l A)(\pa \th^{\b_2} \l^{\a_2} [\tilde{C}(X,\th)]_{\a_2\b_2})(d_{\b_3} \l^{\a_3} [\tilde{E}(X,\th)]^{\b_3}{}_{\a_3})>\nn\\
 \doteq &&  e_1^m k_1^n k_3^p \Big(
  17 \chi_p^\2 \g_n \chi_m^\3 -17 \chi_p^\2 \g_m \chi_n^\3
  +16 \chi_n^\2 \g_p \chi_m^\3-16 \chi_m^\2 \g_p \chi_n^\3
  -5 \chi_r^\2 \g_m\g_n \g_p \chi_r^\3\nn
 \Big)
 \eea

\section{Conclusion}
This work is our continued effort to establish the possible connection, proposed in \cite{Park:2007mc}\cite{Park:2008sg}, between open string quantum effects and the corresponding D-brane geometry.
We believe that the connection, once established, will be one of two main components that may lead to the first-principle derivation of AdS/CFT. The other component would be the conversion of an open string into a closed string discussed in \cite{Park:2001bm}. The connection would not only serve in derivation of AdS/CFT but also dictate change in our notion of geometry (at least the geometry associated with a D-brane) at a fundamental level: geometry would be a secondary effect in the sense it is associated with quantum effects of gauge/open string fields.

A few checks of the proposal were carried out in \cite{Park:2008fp}
and \cite{Park:2009ki}, where divergence cancellation for scattering
of massless states were analyzed at one-loop and two-loop
respectively. Even though the results were consistent with the
proposal, it seems necessary to go to the three-loop in order to see
the involvement of the higher curvature terms. Establishing the
connection, therefore, requires an effective tool for computing
higher loop diagrams. We believe that the pure spinor formulation
potentially provides such a tool.\footnote{Having an effective tool
of computing superstring higher loops will be important for many
applications.} One of the goals of this paper has been to strengthen
our skills in the pure spinor formulation for its future
applications in multi-loop computations. On the other hand, it is in
principle possible that the role of the higher curvature terms may
be revealed even at the two-loop order for scattering of {\em
massive states}. In this paper, we have computed several scattering
amplitudes that involve first massive states at tree and one-loop
level, setting the grounds for two-loop computation. To assure the
correctness of the results, We have carried out the analysis in the
NSR formulation first, and subsequently reproduce the same results
in the pure spinor formulation. The pure spinor computation requires
gauge fixing, which we have discussed in section 3.2. For a general
amplitude it is also necessary to construct the integrated vertex
operator for the massive states, a task that deserves its own work.
It would be interesting to see whether the two-loop
``renormalization" (in the sense of
\cite{Park:2007mc}\cite{Park:2008sg}) of open string theory would
indeed fully verify the physical picture of the open string
loop-induced D-brane geometry.

There are a few other near-future directions. For the last several years, one of the active areas in AdS/CFT has been in matching the anomalous dimensions of certain SYM operators with the energies of the semi-classical configurations of a closed string \cite{Gubser:2002tv}. With the renormalization established, it will be interesting to study how to embed the SYM analysis in a full-fledged open string analysis \cite{progress}.
From our standpoint, it is rather evident that a full-fledged open string analysis should be possible. (The relevance of such an analysis is obvious but there will be more remarks on this below.) The belief is based on a few things that we discuss now. The success of comparing the planar SYM anomalous dimensions with the corresponding semi-classical closed string solitons seems to signify the following. In the past, there were beliefs/attempts to realize a closed string as an open string bound state. Since a closed string would not be in the Fock space, it was expected that the its realization in terms of open string fields would be complex. (The counter vertex operator proposed in \cite{Park:2008fp} can be viewed as one such realization.)
The complexity may be due to the fact that the close string that one attempts to realize is a {\em fundamental} string; if one instead considers a solitonic configuration such as the one in \cite{Gubser:2002tv} (and many others in the related works afterwards), its construction in terms of open string fields may get vastly simplified.\footnote{As a matter fact those kinds of semi-classical configurations may also admit a description by closed string vertex operators as discussed in \cite{Tseytlin:2003ac}. The whole picture, therefore, seems to point towards the generalized open/closed string duality \cite{Park:2001bm}.} Given that SYM is a low energy limit of an open string, the statement can be paraphrased taking the spin chain/AdS correspondence: the semi-classical closed string solitons are more complex than a fundamental closed string, and that shift of the complexity has made the corresponding gauge theory operators simpler.

Getting back to the open string realization of a closed string, it
would be very surprising if the success of SYM anomalous dimensions
could not be extended to the full open string description. Since the
birthplace of AdS/CFT was string theory, it seems not only possible
but also natural that the SYM description of a closed string soliton
admits a full open string description. In addition, there is a much
less understood regime of non-planar SYM. Once the large-N limit is
lifted, one would have to include the entire tower of the open
string massive states anyway. They would contribute to the
``anomalous" dimensions of the massless modes, SYM, by circulating
the loops (and the results should reduce to those of SYM by taking a
low energy limit if so desired). We will report our progress on this
and other related issues in the near future.

\vspace*{1in}
\ni {\bf Acknowledgements:}\\
\ni I thank Deogki Hong for his invitation and hospitality during my visit to Pusan National University. I acknowledge the KITP scholar program and thank Joe Polchinski for his interest in my recent works. My research greatly benefited from my visit to Seoul National University, Kyung Hee University and KIAS: I thank Choonkyu Lee, Nakwoo Kim, Piljin Yi and Kimyeong Lee for their hospitalities and valuable discussions.

\newpage
\renewcommand{\theequation}{A.\arabic{equation}}
\setcounter{equation}{0}
\section*{Appendix A: Useful identities for NSR analysis }
The massless vector four-point amplitudes both at tree and one-loop levels contain the following kinematic factor,
\bea
K=&&-\fr14(st\;\z_1\cdot \z_3\;\z_2\cdot \z_4+
su\;\z_2\cdot \z_3\;\z_1\cdot \z_4
+tu\;\z_1\cdot \z_2\;\z_3\cdot \z_4)\nn\\
&&+\fr12 s(\z_1\cdot k_4\;\z_3\cdot k_2\; \z_2\cdot \z_4
+\z_2\cdot k_3\;\z_4\cdot k_1\; \z_1\cdot \z_3\nn\\
&&\quad\quad +\z_1\cdot k_3\;\z_4\cdot k_2\; \z_2\cdot \z_3
+\z_2\cdot k_4\;\z_3\cdot k_1\; \z_1\cdot \z_4)\nn\\
&&+\fr12 t(\z_2\cdot k_1\;\z_4\cdot k_3\; \z_3\cdot \z_1
+\z_3\cdot k_4\;\z_1\cdot k_2\; \z_2\cdot \z_4\nn\\
&&\quad\quad +\z_2\cdot k_4\;\z_1\cdot k_3\; \z_3\cdot
\z_4
+\z_3\cdot k_1\;\z_4\cdot k_2\; \z_2\cdot \z_1)\nn\\
&& +\fr12 u(\z_1\cdot k_2\;\z_4\cdot k_3\; \z_3\cdot \z_2
+\z_3\cdot k_4\;\z_2\cdot k_1\; \z_1\cdot \z_4\nn\\
&&\quad\quad +\z_1\cdot k_4\;\z_2\cdot k_3\; \z_3\cdot
\z_4
+\z_3\cdot k_2\;\z_4\cdot k_1\; \z_1\cdot \z_2)
\label{K}
\eea
where
\bea
s=-(k_1+k_2)^2 \quad t=-(k_2+k_3)^2 \quad u=-(k_1+k_3)^2
\eea
The following integral relations are useful when evaluating tree level amplitudes:
\bea
\int_0^1 dx\;\fr{1}{x}\;x^{-\a's}(1-x)^{-\a't}
& \Rightarrow & -\a' t \;\nn\\
\int_0^1 dx\;\fr{1}{(1-x)}\;x^{-\a's}(1-x)^{-\a't}
& \Rightarrow& -\a' s \;\nn\\
\int_0^1 dx\;\fr{1}{x(1-x)}\;x^{-\a's}(1-x)^{-\a't}
& \Rightarrow& \a' u \;\nn\\
\int_0^1 dx\;\fr{1}{x^2}\;x^{-\a's}(1-x)^{-\a't} & =& \fr{\a' t\,\a'
u}{1+\a's}
\;\nn\\
\int_0^1 dx\;\fr{1}{(1-x)^2}\;x^{-\a's}(1-x)^{-\a't} & \Rightarrow&
\fr{\a' s\; \a' u}{1+\a' t}
\;\nn\\
\int_0^1 dx\;x^{-\a's}(1-x)^{-\a't} & \Rightarrow& \fr{\a' s\; \a'
t}{1+\a' u}
\;\nn\\
\int_0^1 dx\;\fr{x}{(1-x)^2}\;x^{-\a's}(1-x)^{-\a't} & \Rightarrow&
\left(\a' s+\fr{\a' s\; \a' u}{1+\a' t}\right)
\;\nn\\
\int_0^1 dx\;\fr{1}{(1-x)^2x}\;x^{-\a's}(1-x)^{-\a't} & \Rightarrow&
\left(\a' u+\fr{\a' s\; \a' u}{1+\a' t}\right)
\;\nn\\
\int_0^1 dx\;\fr{1}{(1-x)x^2}\;x^{-\a's}(1-x)^{-\a't} & \Rightarrow&
\left(\a' u+\fr{\a' t\; \a' u}{1+\a' s}\right) \;\nn\\
\int_0^1 dx\;\fr{1}{x^3}\;x^{-\a's}(1-x)^{-\a't} & \Rightarrow&
\fr{(1-\a'u)\;\a' t\; \a' u}{(2+\a' s)(1+\a' s)} \;
\eea
where $\Rightarrow$ indicates the fact that the following factor
\be
\fr{\G(-\a' s)\G(-\a' t)}{\G(1-\a' s-\a' t)}
\ee
has been omitted in the right hand sides.
Here is a summary of our conventions for Jacobi $\vartheta$-functions and several relations used in the NSR analysis. Our conventions for $\vartheta$-functions are basically those of \cite{pol}. For example, the $\eta$-function is given by
\bea
\eta(\t)=\left[\fr{\pa_x \vartheta_{11}(0,\t)}{-2\pi}\right]^{1/3}
\eea
The prime on $\vartheta$
denotes differentiation with respect to the first argument,
\bea
\vartheta'(z,\t)=\frac{\pa }{\pa x}\vartheta(x,\t)
\eea
In section 2, various Riemann identities such as\footnote{these identities can be easily derived by considering one-loop three-point amplitude and using the fact that they vanish due to the non-saturation of the fermionic zero-modes. }
\bea
&&\sum_\n C_\n \vartheta_{ab}(0,\t)^4
S_\n(z_1-z_2)S_\n(z_2-z_3)S_\n(z_3-z_1)=0\nn\\
&& \sum_\n C_\n \vartheta_{ab}(0,\t)^4 S_\n(z_1-z_2)S_\n(z_2-z_1)=0
\label{3ptriemann}
\eea
were used. The appearance of the factor, $\vartheta_{ab}(0,\t)^4$, can be understood as follows.
It is the part of the path-integral that need to be evaluated at some point of the amplitude calculation. They basically come from the $\psi$- and $bc$- kinetic terms \cite{D'Hoker:1988ta}
\bea
\int_{even} D(X\psi bc\b\g)\left[ bc\, \prod e^{ik_i\cdot X^i(z_i)}\right]
e^{iS(X,\psi,b,c,\b,\g)}
\Rightarrow \fr12 \fr{\vartheta_\n(0,\t)^4}{\eta(\t)^{12}}\;
\eea
where $\Rightarrow$ indicates the fact that the usual factor associated with $\cal F$ and some other irrelevant factors have been omitted.
The following Riemann identity was used in several places in the NSR analysis:
\begin{eqnarray}
\sum_{a,b}(-1)^{a+b}\vartheta_{ab}(x)\vartheta_{ab}(y)
\vartheta_{ab}(u)\vartheta_{ab}(v)
&=& 2\vartheta_{11}(x_1)\vartheta_{11}(y_1)\vartheta_{11}(u_1)
\vartheta_{11}(v_1) \label{ri}
\end{eqnarray}
where
\begin{eqnarray}
&& x_1= \fr12 (x+y+u+v),\quad y_1= \fr12 (x+y-u-v)\nn\\
&& u_1= \fr12 (x-y+u-v),\quad v_1= \fr12 (x-y-u+v)
\end{eqnarray}

\renewcommand{\theequation}{B.\arabic{equation}}
\setcounter{equation}{0}
\section*{Appendix B: Useful identities for pure spinor computation }
Our convention for the 32 by 32 gamma matrices are
\[
\G^m=\left(
\begin{array}{cc}
0 & (\g^m)_{\a\b} \\
(\g^m)^{\a\b} & 0 \\
\end{array}
\right)\;\;,\;\;
\]
where $\g^m$'s are the 16 by 16 gamma matrices.
They satisfy\footnote{An extensive list of identities are given here even though only a few of them were used in the actual analysis in the main body and in Appendix C. The list may serve in future computations.}
\bea
&& \G^a\G_a=10,\quad \G^a \G^\m \G_a=-8\G^\m,\quad \G^a \G^{\m\n}
\G_a=6\G^{\m\n}
\nn\\
&&\G^a \G^{\m\n\r} \G_a=-4\G^{\m\n\r},\quad
\G^a\G^{\m\n\r\s} \G_a=2\G^{\m\n\r\s}, \quad
\G^a \G^{\m\n\r\s\d} \G_a=0\nn\\
&&\G^a \G^{\m\n\r\s\d\k} \G_a=-2\G^{\m\n\r\s\d\k},\quad
\G^a \G^{\m\n\r\s\d\k\z} \G_a=4\G^{\m\n\r\s\d\k\z}\nn\\
&&\G^{ab}\G_{ab}=-90,\quad
\G^{ab} \G^\m \G_{ab}=-54\G^\m,\quad \G^{ab} \G
^{\m\n} \G_{ab}=-26\G^{\m\n}\nn\\
&& \G^{ab} \G^{\m\n\r} \G_{ab}=-6\G
^{\m\n\r},\quad\G^{ab}
\G^{\m\n\r\s} \G_{ab}=6\G
^{\m\n\r\s},\quad \G^{ab} \G^{\m\n\r\s\d} \G
_{ab}=10 \G^{\m\n\r\s\d},\quad \nn\\
&&\G^{ab} \G^{\m\n\r\s\d\k} \G
_{ab}=6\G^{\m\n\r\s\d\k}\nn\\
&& \G^{abc}\G_{abc}=-720,\quad
\G^{abc} \G^\m \G_{abc}=288\G^\m,\quad \G^{abc} \G^{\m\n} \G
_{abc}=-48\G^{\m\n}
\nn\\
&& \G^{abc} \G^{\m\n\r} \G_{abc}=-48\G^{\m\n\r},\quad \G^{abc}
\G^{\m\n\r\s} \G_{abc}=48\G^{\m\n\r\s},\quad \G^{abc} \G^{\m\n\r\s\d}
\G_{abc}=0
\nn\\
&& \G^{abc} \G
^{\m\n\r\s\d\k} \G_{abc}=-48\G^{\m\n\r\s\d\k}
\nn\\
&&\G^{abcd}\G_{abcd}=5040,\quad
\G^{abcd} \G^\m \G
_{abcd}=1008\G^\m,\quad \G^{abcd} \G^{\m\n} \G_{abcd}=-336\G^{\m\n}
\nn\\
&&\G^{abcd} \G^{\m\n\r} \G_{abcd}=-336\G^{\m\n\r},\quad \G^{abcd}
\G^{\m\n\r\s} \G
_{abcd}=48\G^{\m\n\r\s}\nn\\
&& \G^{abcd} \G^{\m\n\r\s\d} \G_{abcd}=240 \G
^{\m\n\r\s\d},\quad
\G^{abcd} \G^{\m\n\r\s\d\k} \G_{abcd}=48\G^{\m\n\r\s\d\k}\nn\\
&&
\G^{abcde}\G_{abcde}=6\cdot 5040,\quad
\G^{abcde}\G^\m\G_{abcde}=0,\quad \G^{abcde}\G^{\m\n}\G_{abcde}=-3360\G^{\m\n},
\quad \nn\\
&&\G^{abcde}\G^{\m\n\r}\G_{abcde}=0,\quad
\G^{abcde}\G^{\m\n\r\s}\G_{abcde}=1440\G^{\m\n\r\s}\nn\\
&&
\G^{abcde}\G^{\m\n\r\s\d}\G_{abcde}=0
\eea
and
\bea
&&[\G _{m},\G^{r}]=2\G_{m}{}^{r}\quad,\quad
\{\G _{m},\G^{r}\}=2\d_{m}{}^{r} \nn\\
&& \{\G _{mn},\G^{r}\}=2\G_{mn}{}^{r} \quad,\quad
[\G _{mn},\G
^{r}]=-4\d_{[m}^{r}\G_{n]} \nn\\
&& [\G _{mnp},\G^{r}]=2\G_{mnp}{}^{r}\quad,\quad
\{\G
_{mnp},\G^{r}\}=6\,\d_{[m}^{r}\G_{np]}\nn\\
&& \{\G
_{mnpq},\G^{r}\}=2\G_{mnpq}{}^{r} \quad,\quad
[\G _{mnpq},\G^{r}]=-8\d_{[m}^{r}\G
_{npq]}\nn\\
&& [\G _{mnpqk},\G^{r}]=2\G_{mnpqk}{}^{r} \quad,\quad
\{\G _{mnpqk},\G^{r}\}=10\,\d_{[m}^{r}\G_{npqk]} \nn\\
&& \{\G
_{mn},\G^{rs}\}=2\G_{mn}{}^{rs}-4\d_{[mn]}^{rs} \quad,\quad
[\G _{mn},\G^{rs}]=-8\d_{[m}^{[r} \G_{n]}{}^{s]} \nn\\
&&\{\G
_{mnp}\,,\G^{rs}\}=2\G_{mnp}{}^{rs}-12\d_{[mn}^{rs}\G_{p]}\quad,\quad
[\G _{mnp}\,,\G^{rs}]=12\d_{[m}^{[r}\G_{np]}{}^{s]}
\nn\\
&& \{\G_{mnpq}\,,\G
^{rs}\}=2\G_{mnpq}{}^{rs}-24\d_{[mn}^{rs}\G_{pq]}
\quad,\quad [\G _{mnpq}\,,\G^{rs}]=-16\d_{[m}^{[r}\G
_{npq]}{}^{s]}
\nn\\
&&\{\G_{mnpqk}\,,\G^{rs}\}=2\G_{mnpqk}{}^{rs}-40\d_{[mn}^{rs}\G_{pqk]}
\quad,\quad [\G
_{mnpqk}\,,\G^{rs}]=20\d_{[m}^{[r}\G_{npqk]}{}^{s]}
\nn\\
&&[\G _{mnp},\G
^{rst}]=2\G_{mnp}{}^{rst}-36\;\d_{[mn}^{[rs}\G_{p]}{}^{t]}\,\quad
\{\G _{mnp},\G^{rst}\}=18\d_{[m}^{[r}\G_{np]}{}^{rs]}-12\d_{[mnp]}^{rst}
\nn\\
&&\{\G
_{mnpq}\,,\G^{rst}\}=2\G_{mnpq}{}^{rst}-72\d_{[mn}^{[rs}\G_{pq]}{}^{t]}
\quad\quad\nn\\
&&[\G_{mnpq},\G^{rst}]=-24\d_{[m}^{[r}\G_{npq]}{}^{st]}+48\;\d_{[mnp}^{rst}\G_{q]}
\nn\\
&&\{\G
_{mnpq}\,,\G^{rstu}\}=2\G_{mnpq}{}^{rstu}-144\d_{[mn}^{rs}\G_{pq]}{}^{tu]}
+48\d_{[mnpq]}^{rstu}\quad,\quad\nn\\
&& [\G_{mnpq},\G^{rstu}]=-32\d_{[m}^{[r}\G_{npq}{}^{stu]}
-64\d_{[mnp}^{[rst}\G_{q}{}^{u]}
\nn\\
&& \{\G_{mnpqk},\G
^{rst}\}=30\d_{[m}^{[r}\G_{npqk}{}^{st]}-120\;\d_{[mnp}^{[rst}\G_{qk]}
\nn\\
&&[\G_{mnpqk},\G^{rst}]=2\G
_{mnpqk}{}^{rst}-120\;\d_{[mn}^{[rs}\G_{pqk]}{}^{t]}
\nn\\
&& \{\G_{mnpqk},\G^{rstu}\}=2\G
_{mnpqk}{}^{rstu}-240\;\d_{[mn}^{[rs}\G_{pqk}{}^{tu]}
+240\;\d_{[mnpq}^{[rstu}\G_{k]}\nn\\
&& [\G_{mnpqk},\G^{rstu}]=40\;\d_{[m}^{[r}\G_{npqk}{}^{stu]}
-480\;\d_{[mnp}^{[rst}\G
_{qk]}{}^{u]}\nn\\
&& \{\G_{mnpqk},\G^{rstuw}\}=50\;\d_{[m}^{[r}\G_{npqk}{}^{stuw]}
-1200\;\d_{[mnp}^{[rst}\G_{qk}{}^{uw]}
+240\;\d_{mnpqk}^{rstuw}\nn\\
&& [\G
_{mnpqk},\G^{rstuw}]=2\G_{mnpqk}{}^{rstuw}-400\;\d_{[mn}^{[rs}\G_{pqk}{}^{tuw]}
+1200\;\d_{[mnpq}^{[rstu}\G_{k]}{}^{w]}\nn\\
&& \{\G_{mnpqkl},\G^{r}\}=2\G
_{mnpqkl}{}^{r},\quad
[\G_{mnpqkl},\G^{r}]=-12\;\d_{[m}^{r}\G_{npqkl]}\nn\\
&&\{\G_{mnpqkl},\G^{rs}\}=2\G_{mnpqkl}{}^{rs}-60\;\d_{[mn}^{[rs}\G
_{pqkl]},\quad
[\G_{mnpqkl},\G^{rs}]=-24\;\d_{[m}^{[r}\G_{npqkl]}{}^{s]}
\nn\\
&&\{\G_{mnpqkl},\G^{rst}\}=2\G_{mnpqkl}{}^{rst}-180\;\d_{[mn}^{[rs}\G_{pqkl]}{}^{t]}
\nn\\
&& [\G
_{mnpqkl},\G^{rst}]=-36\d_{[m}^{[r}\G_{npqkl}{}^{st]}
+240\;\d_{[mnp}^{rst}\G_{qkl]}
\nn\\
&& \{\G_{mnpqkl},\G^{rstu}\}=2\G_{mnpqkl}{}^{rstu}-360\;\d_{[mn}^{[rs}\G
_{pqkl}{}^{tu]}
+720\;\d_{[mnpq}^{rstu}\G_{kl]}\nn\\
&& [\G_{mnpqkl},\G^{rstu}]=-48\;\d_{[m}^{[r}\G
_{npqkl}{}^{stu]}
+960\;\d_{[mnp}^{[rst}\G_{qkl]}^{u]}\nn\\
&& \{\G_{mnpqkl},\G^{rstuw}\}=-600\;\d_{[mn}^{[rs}\G_{pqkl}{}^{tuw]}
+3600\;\d_{[mnpq}^{[rstu}\G_{kl]}{}^{w]}\nn\\
&& [\G_{mnpqkl},\G^{rstuw}]=-60\;\d_{[m}^{[r}\G_{npqkl}{}^{stuw]}
+2400\;\d_{[mnp}^{[rst}\G_{qkl}{}^{uw]}
-1440\;\d_{[mnpqk}^{rstuw}\G_{l]}\nn\
\eea
The following relations were used when a transpose of the 32 by 32 gamma matrix was taken,
\bea
(\G^\m)^T&=& \G^0\G^\m \G^0\nn\\
(\G^{\m\n})^T&=&\G^0 \G^{\m\n} \G^0 \nn\\
(\G^{\m\n\r})^T&=&-\G^0 \G^{\m\n\r} \G
^0\nn\\
(\G^{\m\n\r\s})^T&=&-\G^0 \G^{\m\n\r\s} \G^0\nn\\
(\G^{\m_1...\m_5})^T&=&\G^0 \G^{\m_1...\m_5} \G^0
\eea
Some of the identities given in
\cite{Berkovits:2006bk} were used in the computations in section 3.3. To make this paper self-contained we present them below:
\bea
<(\l\g^m\th)(\l\g^n\th)(\l\g^p\th)(\th\g_{ijk}\th)>&=&\fr1{120}\d_{ijk}^{mnp}\nn\\
<(\l\g^{mnp}\th)(\l\g_q\th)(\l\g_t\th)
(\th\g_{ijk}\th)>&=&\fr1{70}\d_{[q}^{[m} \eta_{t][i}\d_j^n\d_{k]}^{p]}
\eea
\bea
&& <(\l\g_{t}\th)(\l\g^{mnp}\th)(\l\g^{qrs}\th)(\th\g_{ijk}\th)>\nn\\
=&& \fr1{8400}\e^{ijkmnpqrst}
+ \fr1{140}\Big(
\d_{t}^{[m} \d_{[i}^{n}\eta^{p][q}\d_j^r\d_{k]}^{s]}
-\d_{t}^{[q} \d_{[i}^{r}\eta^{s][m}\d_j^n\d_{k]}^{p]}
\Big)\nn\\
&&-\fr1{280}\Big(
\eta_{t[i}\eta^{v[q}\d_j^r\eta^{s][m}\d_{k]}^n\d_v^{p]}
-\eta_{t[i}\eta^{v[m}\d_j^n\eta^{p][q}\d_{k]}^r\d_v^{s]}
\Big)
\eea
\bea
&&<(\l\g_{mnpqr}\l)(\l\g^{u}\th)(\th\g^{fgh}\th)(\th\g_{jkl}\th)>
\nn\\
= &&-\fr4{35}\Big(
\d_{[j}^{[m}\d_k^n \d_{l]}^p \d_{[f}^q \d_g^{r]} \d_{h]}^u
+\d_{[f}^{[m}\d_g^n \d_{h]}^p \d_{[j}^q \d_k^{r]} \d_{l]}^u
-\fr12 \d_{[j}^{[m}\d_k^n \eta_{l][f} \d_{g}^p \d_{h]}^q \eta^{r]u}
-\fr12 \d_{[f}^{[m}\d_g^n \eta_{h][j} \d_{k}^p \d_{l]}^q \eta^{r]u}
\Big)\nn\\
&&-\fr1{1050} \e^{mnpqr}{}_{abcde}\Big(
\d_{[j}^{[a}\d_k^b \d_{l]}^c \d_{[f}^d \d_g^{e]} \d_{h]}^u
+\d_{[f}^{[a}\d_g^b \d_{h]}^c \d_{[j}^d \d_k^{e]} \d_{l]}^u\nn\\
&&\hspace{1.5in}
-\fr12 \d_{[j}^{[a}\d_k^b \eta_{l][f} \d_{g}^c \d_{h]}^d \eta^{e]u}
-\fr12 \d_{[f}^{[a}\d_g^b \eta_{h][j} \d_{k}^c \d_{l]}^d \eta^{e]u}
\Big)
\eea
\bea
&& <(\l\g^{mnpqr}\th)(\l\g_{stu}\th)(\l\g_v\th)(\th\g_{fgh}\th)>\nn\\
=&&
\fr1{35}\eta^{v[m}\d_{[s}^{n}\d_t^p \eta_{u][f}\d_g^q\d_{h]}^{r]}
-\fr2{35}\d_{[s}^{[m}\d_{t}^{n}\d_{u]}^{p}
\d_{[f}^{q}\d_{g}^{r]}\d_{h]}^{v}\nn\\
&&
+\fr1{120}\e^{mnpqr}{}_{abcde}\Big(
\fr1{35}\eta^{v[a}\d_{[s}^{b}\d_t^c \eta_{u][f}\d_g^d\d_{h]}^{e]}
-\fr2{35}\d_{[s}^{[a}\d_{t}^{b}\d_{u]}^{c}
\d_{[f}^{d}\d_{g}^{e]}\d_{h]}^{v}
\Big)
\eea
\bea
&& <(\l\g^{mnpqr}\th)(\l\g_{d}\th)(\l\g_e\th)(\th\g_{fgh}\th)>
=
-\fr1{42}\d_{abcde}^{mnpqr}
-\fr1{5040}\e^{mnpqr}{}_{defgh}
\eea
\renewcommand{\theequation}{C.\arabic{equation}}
\setcounter{equation}{0}
\section*{Appendix C: Contribution of $N^{mn}\l^\a G(X,\th)_{mn\a}$ to AAb}
In section 3.3, we computed the contribution of $F$ in \rf{Vschematic} to AAB-amplitude at tree level.
Here we present the contributions of the terms that contain,
\bea
E(X,\th)^\b{}_\a &=& \fr12 ( \g^{mnpq})^\b{}_\a \pa_m b_{npq} \nn\\
G(X,\th)_{mn\a} &=& \fr18 \left[
3 ( \g_{st}\g^l \th)_\a \pa_m \pa_l \,b^{st}{}_n
+\fr37 ( \g_{qm}\g_{st}\g^l \th)_\a\, \pa^q \pa_l\, b^{st}{}_n
\right]
\eea
Following the steps similar to those of section 3.3, one can show
\bea
&&<\l A^\1\,\l A^\2\,d_\b \l^\a [E(X,\th)^\3]^\b{}_\a>\nn\\
=&& \fr{i}{96\cdot 576}\z_1^m \z_2^{p'}e_{3npq}k_2^s k_2^{q'}k_3^m\nn\\
&&\Big[
<(\l \g^{m_2} \g^{mnpq} \l)(\l \g^{m_1}\th)(\th \g_{m_2}{}^{sn_2}\th)
(\th \g_{n_2}{}^{p'q'}\th) >\nn\\
&& -2<(\l \g^{mnpq} \g^{m_2sn_2} \th)(\l \g^{m_1}\th)(\l \g_{m_2}\th)
(\th \g_{n_2}{}^{p'q'}\th) >\nn\\
&& -2<(\l \g^{mnpq} \g^{n_2p'q'} \th)(\l \g^{m_1}\th)(\l \g_{m_2}\th)
(\th \g_{m_2}{}^{sn_2}\th) >
\label{E}
\Big]+(1\leftrightarrow 2)
\eea
The term of $(1\leftrightarrow 2)$ entirely vanishes according to Mathematica computation; so does the first term in \rf{E}.
Using the gamma matrix identities given in Appendix B, the second term in \rf{E} can be rewritten as
\bea
&& \z_1^m \z_2^{p'}e_{3npq}k_2^s k_2^{q'}k_3^m
<(\l \g^{mnpq} \g^{m_2sn_2} \th)(\l \g^{m_1}\th)(\l \g_{m_2}\th)
(\th \g_{n_2}{}^{p'q'}\th) >\nn\\
=&& \z_1^m \z_2^{p'}e_{3npq}k_2^s k_2^{q'}k_3^m\nn\\
&&\Big(
<(\l \g_{mnpq}\g^{m_2} \g^{sn_2} \th)(\l \g^{m_1}\th)(\l \g_{m_2}\th)
(\th \g_{n_2}{}^{p'q'}\th) >\nn\\
&&-<(\l \g_{mnpqn_2} \th)(\l \g^{m_1}\th)(\l \g_{s}\th)
(\th \g_{n_2}{}^{p'q'}\th) >\nn\\
&&+4<(\l \d_{[m}^{n_2} \g_{npq]} \th)(\l \g^{m_1}\th)(\l \g_{s}\th)
(\th \g_{n_2}{}^{p'q'}\th) >\nn\\
&&+<(\l \g_{mnpqs} \th)(\l \g^{m_1}\th)(\l \g_{s}\th)
(\th \g_{n_2}{}^{p'q'}\th) >\nn\\
&&-4<(\l \d_{[m}^{s} \g_{npq]} \th)(\l \g^{m_1}\th)(\l \g_{n_2}\th)
(\th \g_{n_2}{}^{p'q'}\th) >
\Big)
\label{E2nd}
\eea
It turns out that only the first term yields a non-vanishing result. The first term can be rewritten as
\bea
&& \z_1^m \z_2^{p'}e_{3npq}k_2^s k_2^{q'}k_3^m
<(\l \g_{mnpq}\g^{m_2} \g^{sn_2} \th)(\l \g^{m_1}\th)(\l \g_{m_2}\th)
(\th \g_{n_2}{}^{p'q'}\th) >\nn\\
=&& 6\,\z_1^m \z_2^{p'}e_{3npq}k_2^s k_2^{q'}k_3^m
\Big(<(\l \g_{mpqsn_2} \th)(\l \g^{m_1}\th)(\l \g_{n}\th)
(\th \g_{n_2}{}^{p'q'}\th) >\nn\\
&&\hspace{1.3in} -6<(\l \d_{[mp}^{sn_2} \g_{q]} \th)(\l \g^{m_1}\th)(\l \g_{n}\th)
(\th \g_{n_2}{}^{p'q'}\th) >\nn\\
&&\hspace{1.3in} +6<(\l \d_{[m}^{s} \g_{pq]}{}^{n_2} \th)(\l \g^{m_1}\th)(\l \g_{n}\th)
(\th \g_{n_2}{}^{p'q'}\th) >
\Big)\nn\\
&& -2\,\z_1^m \z_2^{p'}e_{3npq}k_2^s k_2^{q'}k_3^m
\Big(<(\l \g_{npqsn_2} \th)(\l \g^{m_1}\th)(\l \g_{m}\th)
(\th \g_{n_2}{}^{p'q'}\th) >\nn\\
&&\hspace{1.3in} -6<(\l \d_{[np}^{sn_2} \g_{q]} \th)(\l \g^{m_1}\th)(\l \g_{m}\th)
(\th \g_{n_2}{}^{p'q'}\th) >\nn\\
&&\hspace{1.3in} +6<(\l \d_{[n}^{s} \g_{pq]}{}^{n_2} \th)(\l \g^{m_1}\th)(\l \g_{m}\th)
(\th \g_{n_2}{}^{p'q'}\th) >
\Big) \label{E2nd_1st}
\eea
Each of the term in \rf{E2nd_1st} is analyzed with the following
results:
\[
\z_1^m \z_2^{p'}e_{3npq}k_2^s k_2^{q'}k_3^m
<(\l \g_{mpqsn_2} \th)(\l \g^{m_1}\th)(\l \g_{n}\th)
(\th \g_{n_2}{}^{p'q'}\th) >
= -\fr1{420}
\z_1^{\m} \z_2^{\n}k_2^\r e_{3\m\n\r}\,k_2\cdot k_3
\]
\[
\z_1^m \z_2^{p'}e_{3npq}k_2^s k_2^{q'}k_3^m
<(\l \d_{[mp}^{sn_2} \g_{q]} \th)(\l \g^{m_1}\th)(\l \g_{n}\th)
(\th \g_{n_2}{}^{p'q'}\th) >
= -\fr1{2160}\,\z_1^{\m} \z_2^{\n}k_2^\r e_{3\m\n\r}\,k_2\cdot k_3
\]
\[
\z_1^m \z_2^{p'}e_{3npq}k_2^s k_2^{q'}k_3^m <(\l \d_{[m}^{s}
\g_{pq]}{}^{n_2} \th)(\l \g^{m_1}\th)(\l \g_{n}\th)
(\th \g_{n_2}{}^{p'q'}\th) >
=-\fr1{1512}\, \z_1^{\m} \z_2^{\n}k_2^\r e_{3\m\n\r}\,k_2\cdot k_3
\]
\[
\z_1^m \z_2^{p'}e_{3npq}k_2^s k_2^{q'}k_3^m <(\l \g_{npqsn_2} \th)
(\l \g^{m_1}\th)(\l \g_{m}\th)(\th \g_{n_2}{}^{p'q'}\th) >
=-\fr1{140 } \z_1^{\m} \z_2^{\n}k_2^\r e_{3\m\n\r}\,k_2\cdot k_3
\]
\[
\z_1^m \z_2^{p'}e_{3npq}k_2^s k_2^{q'}k_3^m
<(\l \d_{[np}^{sn_2} \g_{q]} \th)(\l \g^{m_1}\th)(\l \g_{m}\th)
(\th \g_{n_2}{}^{p'q'}\th) >
= -\fr1{720 } \z_1^{\m} \z_2^{\n}k_2^\r e_{3\m\n\r}\,k_2\cdot k_3
\]
\[
\z_1^m \z_2^{p'}e_{3npq}k_2^s k_2^{q'}k_3^m
<(\l \d_{[n}^{s} \g_{pq]}{}^{n_2} \th)(\l \g^{m_1}\th
(\l \g_{m}\th)(\th \g_{n_2}{}^{p'q'}\th)
=-\fr1{180 } \z_1^{\m} \z_2^{\n}k_2^\r e_{3\m\n\r}\,k_2\cdot k_3
\]
Combining these, the first term of \rf{E2nd} becomes
\[
\z_1^m \z_2^{p'}e_{3npq}k_2^s k_2^{q'}k_3^m
<(\l \g_{mnpq}\g^{m_2} \g^{sn_2} \th)(\l \g^{m_1}\th)(\l \g_{m_2}\th)
(\th \g_{n_2}{}^{p'q'}\th) >
=\fr3{70}
\z_1^{\m} \z_2^{\n}k_2^\r e_{3\m\n\r}\,k_2\cdot k_3
\]
The third term in \rf{E}
\bea
&& \z_1^m \z_2^{p'}e_{3npq}k_2^s k_2^{q'}k_3^m
<(\l \g^{mnpq} \g^{m_2sn_2} \th)(\l \g^{m_1}\th)(\l \g_{m_2}\th)
(\th \g_{n_2}{}^{p'q'}\th) >\nn\\
=&& \z_1^m \z_2^{p'}e_{3npq}k_2^s k_2^{q'}k_3^m\nn\\
&&\Big(
-\fr13<(\l \g_{mnpqn_2p'q'} \th)(\l \g^{m_1}\th)(\l \g_{m_2}\th)
(\th \g_{m_2}{}^{sm_2}\th) >\nn\\
&&-36<(\l \d_{[mn}^{n_2p'} \g_{pq]}{}^{q'} \th)(\l \g^{m_1}\th)(\l \g_{m_2}\th)
(\th \g_{m_2}{}^{sn_2}\th) >\nn\\
&&-12<(\l \d_{[m}^{n_2} \g_{npq]}{}^{p'q'} \th)(\l \g^{m_1}\th)(\l \g_{m_2}\th)
(\th \g_{m_2}{}^{sn_2}\th) >\nn\\
&&+24<(\l \d_{[mnp}^{n_2p'q'} \g_{q]} \th)(\l \g^{m_1}\th)(\l \g_{m_2}\th)
(\th \g_{m_2}{}^{sn_2}\th) >
\Big)
\label{E3rd}
\eea
After some algebra, it was found that all of the four terms in \rf{E3rd} vanish. As announced above \rf{df}, the overall result of the tree amplitude shows that the kinematic factor is the same as \rf{df} up to a numerical constant. The amplitude from the fourth term of \rf{Vschematic} is
\bea
&&<(\l A^\1)\,(\l A^\2)\,N^{mn}\l^\a G(X,\th)_{mn\a}^\3>\\
= &&<\fr12 \fr{(\g_{mn}\l)^{\a_1}}{z_3-z_1}A_{\a_1}^{\1}\;
(\l A^{\2})\l^\a G(X,\th)_{mn\a}^\3 >
+<(\l A^{\1})\fr12 \fr{(\g_{mn}\l)^{\a_2}}{z_3-z_2}A_{\a_2}^{\2}\;
\l^\a G(X,\th)_{mn\a}^\3 > \nn
\eea
where the equality is obtained by applying the operator product expansion of $\l$ and $N^{mn}$. As in section 3, we choose $x_1=\infty, x_2=1, x_3=0$; the first term vanishes. Explicitly substituting the expression for $G(X,\th)_{mn\a} $, one gets
\bea
=&& \fr1{8}<(\l A^\1)(\l \g_{mn}A^\2)
\Big[3(\l \g_{st}\g^l \th)k_3^l k_3^m e_3^{stn}
+\fr37 (\l \g_{qm}\g_{st}\g^l \th)k_3^q k_3^l e_3^{stn}
\Big]>\nn\\ \label{Gmother}
\eea
Upon substituting \rf{SYM} in the equation above,
the first term of \rf{Gmother} yields
\bea
&&\fr1{8}<(\l A^\1)(\l \g_{mn}A^\2)
(\l \g_{st}\g^l \th)k_3^l k_3^m e_3^{stn}>\nn\\
=&&-\fr{3i}{16^2}\; \z_1^{m_1}\z_2^{n_2}e_3^{stn}k_2^{m_2}k_3^l k_3^m
<(\l \g^{st}\g^{l}\th)(\l\g^{m_1}\th)
(\l \g^{mn}\g^{p_2}\th)(\th \g^{m_2n_2p_2}\th)>\nn\\
&&-\fr{3i}{16^2}\; \z_1^{n_1}\z_2^{m_2}e_3^{stn}k_1^{m_1}k_3^l k_3^m
<(\l \g^{st}\g^{l}\th)(\l\g^{p_1}\th)
(\l \g^{mn}\g^{m_2}\th)(\th \g^{m_1n_1p_1}\th)>
\label{G}\nn\\
\eea
Using one of the gamma matrix identities, the first term of \rf{G} (omitting the overall numerical coefficient) can be re-expressed as
\bea
&& \z_1^{m_1}\z_2^{n_2}e_3^{stn}k_2^{m_2}k_3^l k_3^m
<(\l \g^{st}\g^{l}\th)(\l\g^{m_1}\th)
(\l \g^{mn}\g^{p_2}\th)(\th \g^{m_2n_2p_2}\th)>\nn\\
=&& \z_1^{m_1}\z_2^{n_2}e_3^{stn}k_2^{m_2}k_3^l k_3^m
\Big(
<(\l \g^{stl}\th)(\l\g^{m_1}\th)
(\l \g^{mnp_2}\th)(\th \g^{m_2n_2p_2}\th)>\nn\\
&& \hspace{1.4in}-<(\l \g^{stl}\th)(\l\g^{m_1}\th)
(\l \g^{n}\th)(\th \g^{m_2n_2m}\th)>\nn\\
&& \hspace{1.4in}+<(\l \g^{stl}\th)(\l\g^{m_1}\th)
(\l \g^{m}\th)(\th \g^{m_2n_2n}\th)>
\Big)
\label{G1st}
\eea
Computation based on the Mathematica package, Gamma.m, yields for the first term of \rf{G}
\bea
= -\fr1{3780} \z_1^\m \z_2^\n k_2^\r e_{3\m\n\r}\, k_3^2
\eea
Similarly, the second term in \rf{G} gives
\bea
&&\z_1^{n_1}\z_2^{m_2}e_3^{stn}k_1^{m_1}k_3^l k_3^m
<(\l \g^{st}\g^{l}\th)(\l\g^{p_1}\th)
(\l \g^{mn}\g^{m_2}\th)(\th \g^{m_1n_1p_1}\th)>\nn\\
=&&-\fr{11}{1260} \z_1^\m \z_2^\n k_1^\r e_{3\m\n\r}\, k_3^2
\eea
These results combine implies for the first term of \rf{Gmother} that
\bea
&&\fr1{8}<(\l A^\1)(\l \g_{mn}A^\2)
(\l \g_{st}\g^l \th)k_3^l k_3^m e_3^{stn}>
\doteq i\,\z_1^\m \z_2^\n k_2^\r e_{3\m\n\r}\, k_3^2
\eea
Following the similar steps, one can show that (after some tedious algebra) the second term of \rf{Gmother} yields
\bea
&& \fr1{8} \fr37<(\l A^\1)(\l \g_{mn}A^\2)
(\l \g_{qm}\g_{st}\g^l \th)k_3^q k_3^l e^{stn}>\nn\\
\doteq &&i\,\z_1^\m \z_2^\n k_1^\r e_{3\m\n\r}\, k_3^2
\eea
where $\doteq$ indicates that the overall numerical coefficient is not recorded precisely.

\newpage

\end{document}